\documentclass{article}

\usepackage{arxiv}

\usepackage[utf8]{inputenc} 
\usepackage[T1]{fontenc}    
\usepackage{url}            
\usepackage{booktabs}       
\usepackage{amsfonts}       
\usepackage{nicefrac}       
\usepackage{microtype}      

\usepackage{amsfonts}
\usepackage[misc]{ifsym}
\usepackage{amsmath}
\usepackage{color}
\usepackage[dvipsnames,table,xcdraw]{xcolor}
\usepackage{chngcntr}
\usepackage{booktabs}
\usepackage{amsthm}
\usepackage[colorlinks]{hyperref}
\usepackage{cleveref}
\usepackage{multirow}
\usepackage{mathtools}
\usepackage{thmtools}
\usepackage{enumitem}
\usepackage[ruled,linesnumbered,noend]{algorithm2e}
\usepackage{lipsum,graphicx,subcaption}

\newtheorem{definition}{Definition}
\newtheorem{theorem}{Theorem}
\newtheorem{lemma}{Lemma}

\newtheorem{proposition}{Proposition}

\newtheorem{eg}{Example}
\newtheorem*{proof*}{Proof}
\newtheorem*{proof2*}{Proof of Theorem 2}
\newtheorem*{pfs*}{Proof Sketch}
\usepackage[noend]{algorithmic}
\newcommand{\norm}[1]{\left\lVert#1\right\rVert}
\DeclarePairedDelimiter{\ceil}{\lceil}{\rceil}

\DeclareMathOperator*{\argmin}{arg\,min}

\title{FourierSAT: A Fourier Expansion-Based Algebraic Framework for Solving Hybrid Boolean Constraints \thanks{The author list has been sorted alphabetically by last name; this
should not be used to determine the extent of authors’ contributions.}}

\author{Anastasios Kyrillidis, Anshumali Shrivastava, Moshe Y. Vardi, Zhiwei Zhang \thanks{Corresponding author: Zhiwei Zhang.}\\
Rice University, Houston, TX, USA \\
\{anastasios, anshumali, vardi, zhiwei\}@rice.edu
}

\title{FourierSAT: A Fourier Expansion-Based Algebraic Framework for Solving Hybrid Boolean Constraints \thanks{The author list has been sorted alphabetically by last name; this
should not be used to determine the extent of authors’ contributions.}}

\begin{document}
\maketitle
 
\setcounter{secnumdepth}{1} 

%

\begin{abstract}
The Boolean SATisfiability problem (SAT) is of central importance in computer science. 
Although SAT is known to be NP-complete, progress on the engineering side---especially that of Conflict-Driven Clause Learning (CDCL) and Local Search SAT solvers---has been remarkable. 
Yet, while SAT solvers, aimed at solving industrial-scale benchmarks in Conjunctive Normal Form
(\texttt{CNF}), have become quite mature, SAT solvers that are effective on other types of constraints (e.g., cardinality constraints and \texttt{XOR}s) are less well-studied; a general approach to handling non-\texttt{CNF} constraints is still lacking.
 In addition, previous work indicated that for specific classes of benchmarks, the running time of extant SAT solvers depends heavily on properties of the formula and details of encoding, instead of the scale of the benchmarks, which adds uncertainty to expectations of running time.

To address the issues above, we design \texttt{FourierSAT} \footnote{The tool is available at \url{https://github.com/vardigroup/FourierSAT}}, an incomplete SAT solver based on Fourier analysis of Boolean functions, a technique to represent Boolean functions by multilinear polynomials. 
By such a reduction to continuous optimization, we propose an algebraic framework for solving systems consisting of different types of constraints. 
The idea is to leverage gradient information to guide the search process in the direction of local improvements. Empirical results demonstrate that \texttt{FourierSAT} is more robust than other solvers on certain classes of benchmarks. 
\end{abstract}
\section{Introduction}
 Constraint satisfaction problems (CSPs) are fundamental in mathematics, physics, and computer science.  The Boolean SATisfiability problem (SAT) is a special class of CSPs, where each variable takes value from the binary set $\{\texttt{True}, \texttt{False}\}$.  Solving SAT efficiently is of utmost significance in computer science, both from a theoretical and a practical perspective.

As a special case of SAT, conjunctive normal forms (\texttt{CNF}s) are a conjunction (\texttt{and}-ing) of disjunctions (\texttt{or}-ing) of literals. 
Despite the NP-completeness of \texttt{CNF}-SAT, there has been a lot of progress on the engineering side of \texttt{CNF}-SAT solvers. Mainstream  SAT solvers can be classified into complete and incomplete ones: A complete SAT solver will return a solution if there exists one or prove unsatisfiability if no solution exists,  while an incomplete algorithm is not guaranteed to find a satisfying assignment. Clearly, an incomplete algorithm cannot prove unsatisfiability.

Most modern complete SAT solvers are based on the the Conflict-Driven Clause Learning (CDCL) algorithm \cite{marques1999grasp}, an evolution of the backtracking Davis-Putnam-Logemann-Loveland (DPLL) algorithm \cite{davis1960computing,davis1962machine}. 
The main techniques used in CDCL solvers include clause-based learning of conflicts, random restarts, heuristic variable selection, and effective constraint-propagation data structures \cite{SAT-survey}. Examples of highly efficient complete SAT solvers include 
MiniSat \cite{minisat},  BerkMin \cite{BerkMin}, 
PicoSAT \cite{PicoSAT}, Lingeling \cite{lingeling}, Glusose \cite{glucose} and MapleSAT \cite{maplesat}.
Overall, CDCL-based SAT solvers constitute a huge success for SAT problems, and have been dominating in the research of SAT solving.

Local search techniques are mostly used in incomplete SAT solvers. The number of unsatisfied clauses is often regarded as the objective function. Local search algorithms mainly include greedy local search (GSAT) \cite{GSAT} and random walk GSAT (WSAT) \cite{walksat}. During the main loop, GSAT repeatedly checks
the current assignments neighbors and selects a new assignment to maximize
the number of satisfied clauses. In contrast, WSAT
randomly selects a variable in an unsatisfied clause and inverts
its value \cite{SAT-survey}.
On top of these basic algorithms, several heuristics for variable selection have been proposed, such as NSAT \cite{Nsat}, Sparrow \cite{sparrow}, and ProbSAT \cite{probSAT}. While practical local search SAT solvers could be slower than CDCL  solvers, local search techniques are still useful for solving a certain class of benchmarks, such as hard random formulas and MaxSAT. Local search algorithms can also have nice theoretical properties \cite{An-improved-deterministic-local-search-algorithm}.

Non-\texttt{CNF} constraints are playing important roles in theoretical computer science and other engineering areas, \emph{e.g.}, \texttt{XOR} constraints in cryptography \cite{Biclique-Cryptanalysis-of-the-Full-AES} as well as cardinality constraints (\texttt{CARD}) and Not-all-equal (\texttt{NAE}) constraints in discrete optimization \cite{Graph-coloring-with-cardinality-constraints,nae-coloring}. 
The combination of different types of constraints enhances the expressive power of Boolean formulas; e.g., \texttt{CARD}-\texttt{XOR} is necessary and sufficient for maximum likelihood decoding (MLD) \cite{MLD}, one of the most crucial problems in coding theory. 
Nevertheless, compared to that of \texttt{CNF}-SAT solving, efficient SAT solvers that can handle non-\texttt{CNF} constraints are less well studied. 

One way to deal with non-\texttt{CNF} constraints is to encode them in \texttt{CNF}. 
However, different encodings differ from the size, the ability to detect inconsistencies by unit propagation (arc consistency) and solution density \cite{encoding-handbook-of-satisfiability}. 
It is generally observed that the running time of SAT solvers relies heavily on the detail of encodings. 
E.g., CDCL solvers benefit from arc-consistency \cite{Decomposing-Global-Grammar-Constraints}, while local search solvers prefer short chains of variable  dependencies \cite{Exploiting-Variable-Dependency-in-Local-Search}.  
Finding a best encoding for a solver usually requires considerable testing and comparison \cite{Exploiting-Cardinality-Encodings-in-Parallel-Maximum-Satisfiability}.

Another way is to extend the existing SAT solvers to adapt to non-\texttt{CNF} clauses. 
Work on this line includes Cryptominisat \cite{cmspaper} for \texttt{\texttt{CNF}}-\texttt{XOR}, MiniCARD \cite{minicard} for \texttt{\texttt{CNF}}-\texttt{CARD}, Pueblo \cite{Pueblo} and RoundingSAT \cite{roundingsat} for \texttt{\texttt{CNF}}-Pseudo Boolean and MonoSAT \cite{monosat2015} for handling \texttt{\texttt{CNF}}-graph properties formulas. 
Such specialized extensions, however, often require different techniques for different types of constraints. 
Meanwhile, general ideas for solving hybrid constraints uniformly are still lacking.

\smallskip
\noindent \textbf{Contributions.} 
The primary contribution of this work is the design of a novel algebraic framework as well as a versatile, robust incomplete SAT solver---\texttt{FourierSAT}---for solving hybrid Boolean constraints.

The main technique we used in our method is the Walsh-Fourier transform (Fourier transform, for short) on Boolean functions \cite{O'Donnell:2014:ABF:2683783}. By transforming Boolean functions into ``nice'' polynomials, numerous properties can be analyzed mathematically. 
Besides the success of Fourier transform in theoretical computer science \cite{Constant-depth-circuits-Fourier-transform-and-learnability,General-Bounds-on-Satisfiability-Thresholds-for-Random-CSPs}, recently, this tool has also found surprising uses in algorithm design  \cite{Beyond-Parity-Constraints,Variable-Elimination-in-the-Fourier-Domain}.
In \cite{A-two-phase-algorithm-for-solving-a-class-of-hard-satisfiability}, the authors used a polynomial representation to design a SAT solver. 
However, their solver was still DPLL-based and mathematical properties of polynomials were not fully exploited. 
More algorithmic uses of this technique are waiting to be discovered. 

To our best knowledge, this paper will be the first algorithmic work to study Fourier expansions of Boolean functions in the real domain instead of only Boolean domain. 
After this relaxation, we find Fourier expansions well-behaved in the real domain. 
Thus, we manage to reduce satisfiability of Boolean constraints to continuous optimization and apply gradient-based methods. 
One of the attractive properties of our method is, different types of constraints are handled uniformly---we no longer design specific methods for each type of constraints. 
Moreover, as long as the Fourier expansions of a new type of constraints are known, our solver can be extended trivially. 
In addition, we explain the intuition of why doing continuous optimization for SAT is better than doing discrete local search.

Furthermore, our study on the local landscape of Fourier expansions reveals that the existing of saddle points is an obstacle for us to design algorithms with theoretical guarantees. 
Previous research shows that gradient-based algorithms are in particular susceptible to saddle point problems \cite{Escaping-From-Saddle-Points}. 
Although the study of \cite{Stochastic_Gradient_Descent_Escapes_Saddle_Points_Efficiently,Gradient-Descent-Converges-to-Minimizers} indicate that stochastic gradient descent with random noise is enough to escape saddle points, strict saddle property, which is not valid in our case, is assumed in the work above. 
Therefore, we design specialized algorithms for optimizing Fourier expansions.

Finally, we demonstrate, by experimental results, that for certain class of hybrid formulas, \texttt{FourierSAT} is more robust compared to existing tools. 

We believe that the natural reduction from SAT to continuous optimization, combined with state-of-the-art constrained-continuous optimization techniques, opens a new line of research on SAT solving. 

\section{Notations and Preliminaries}
\subsection{Boolean Formulas and Clauses}
	Let $x=(x_1,...,x_n)$ be a sequence of $n$ Boolean variables. A Boolean function $f(x)$ is a mapping from a Boolean vector $\{\texttt{True,False}\}^n$ to $\{\texttt{True},\texttt{False}\}$.
	A vector $a\in\{\texttt{True},\texttt{False}\}^n$ is called an assignment and $f(a)$ denotes the value of $f$ on $a$. 
	A literal is either a variable $x_i$ or its negation $\neg x_i$. A formula $f=c_1\wedge c_2\wedge \dots \wedge c_m$ is the conjunction of $m$ Boolean functions, where each $c_i$ is called a clause and belongs to a type from the list below:
	\begin{itemize}
	    \item \texttt{\texttt{CNF}:} A \texttt{CNF} clause is a disjunction of elements in a literal set, which is satisfied when at least one literal is true. E.g., $(x_1\vee x_2 \vee x_3)$.
	    \item \texttt{CARD:} Given a set $L$ of variables and an integer $k\in[n]$, a cardinality constraint $D^{\ge k}(L)$ (resp. $D^{\le k}(L)$) requires the number of \texttt{True} variables to be at least (resp. most) $k$. E.g., $D^{\ge 2}(\{x_1,x_2,x_3\})$: $x_1+x_2+x_3\ge 2$.
	    \item \texttt{XOR:} An XOR clause indicates the parity of the number of variables assigned to be \texttt{True}, which can be computed by addition on $\texttt{GF}(2)$. E.g., $x_1\oplus x_2\oplus x_3$.
	    \item \texttt{NAE:} A Not-all-equal (\texttt{NAE}) constraint is satisfied when not all the variables have the same value. E.g., $\texttt{NAE}(1,1,1,1,0)=1$; $\texttt{NAE}(0,0,0,0,0)=0$
	\end{itemize}
	
	 Let the set of clauses of $f$ be $C(f)$. Let $m=|C(f)|$ and $n$ be the number of variables of $f$. A solution of $f$ is an assignment that satisfies all the clauses in $C(f)$. We aim to design a framework to find a solution of $f$.

	\subsection{Fourier Expansion of a Boolean Function}
	We define a Boolean function by $f: \{\pm 1\}^n \to \{\pm 1\}$. One point which might be  counter-intuitive is  that $-1$ is used to stand for \texttt{True} and $+1$ for \texttt{False}.  
	An assignment now is a vector $a\in\{ \pm 1\}^n$.
	
	Fourier expansion is a method for transforming a Boolean function into a multilinear polynomial. 
	The following theorem shows that any function defined on a Boolean hyper-cube has an equivalent Fourier representation.  

	\begin{theorem}\label{FourierTransformation}(Walsh-Fourier Transform) Given a function $f: \{\pm 1\}^n \to \mathbb{R}$, there is a unique way of expressing $f$ as a multilinear polynomial, with at most $2^n$ terms in $S$, where each term corresponds to one subset of $[n]$, according to:	    $$
		f(X) = \sum_{S\subseteq [n]} \left( \widehat{f}(S) \cdot \prod_{i\in S}x_i \right),
		$$ 
		where $\widehat{f}(S)$'s$\in \mathbb{R}$ are called the Fourier coefficients, given $S$, and computed as:
		\begin{small}
		\begin{equation}\nonumber
		    \begin{split}
		        \widehat{f}(S) &= \underset{x\sim \{\pm 1\}^n}{\mathbb{E}} \left[f(x) \cdot \prod_{i\in S}x_i \right] 
		        = \frac{1}{2^n} \!\!\! \sum_{x\in \{\pm 1\}^n} \left(f(x) \cdot \prod_{i\in S}x_i\right)
		    \end{split}
		\end{equation}
		\end{small}
		where $x\sim \{\pm 1\}^n$ indicates $x$ is chosen uniformly from $\{\pm 1\}^n$.
		The polynomial is referred  as the \textbf{Fourier expansion} of $f$. 
	\end{theorem}
	
	\begin{table}
		\centering
		\begin{tabular}{c c c}
			\toprule 
			Clause & & Fourier Expansion \\
			\cmidrule{1-1} \cmidrule{3-3}
			$x_1\vee x_2$ & & $ -\frac{1}{2}+\frac{1}{2}x_1+\frac{1}{2}x_2+\frac{1}{2}x_1x_2$ \vspace{0.2cm}   \\ 
			$D^{\ge2}({x_1,x_2,x_3})$ & & $\frac{1}{2}x_1+\frac{1}{2}x_2+\frac{1}{2}x_3-\frac{1}{2}x_1x_2x_3$  \vspace{0.2cm}\\ 
			$x_1\oplus x_2\oplus x_3$ & & $x_1x_2x_3$ \vspace{0.2cm} \\ 
			$\texttt{NAE}(x_1,x_2,x_3)$ & & $-\frac{1}{2}+\frac{1}{2}x_1x_2+\frac{1}{2}x_2x_3+\frac{1}{2}x_1x_3$  \\ 
			\bottomrule
		\end{tabular} 
		\caption{Examples of Fourier Expansion on different  clauses}
		\label{ex_FE}
	\end{table}
	
	Table \ref{ex_FE} shows some examples of Fourier expansions.
	
\section{A Reduction from SAT to Continuous Optimization}

	In this section, we reduce finding a solution of a Boolean formula to finding a minimum of a continuous multivariate polynomial, based on Fourier expansion. Due to the space limit, we delay most of the proofs to the supplemental material.

	\subsection{A Reduction to a Minimization Problem}
		Our basic idea is to do continuous optimization on Fourier expansions of formulas. However, in general, computing the Fourier coefficients of a Boolean function is nontrivial and often harder than SAT itself (for an example, see Proposition \ref{proposition:pseudo}). Even if we are able to get the Fourier expansion of a Boolean formula, the polynomial can have high degree---and as a result, exponentially many terms---making it hard to store and evaluate. 

	\begin{proposition}
		\label{proposition:pseudo} Computing the Fourier Expansion of a pseudo-Boolean constraint (a generalization of a cardinality constraint, e.g., $3x_1+2x_2+7(\neg x_3)\ge 0$, $x\in \{\pm 1\}^3$) is \#P-hard.
	\end{proposition}
	
	Instead of computing Fourier coefficients of a monolithic logic formula, we take advantage of factoring, constructing a polynomial for a formula by the Fourier expansions of its clauses. Excitingly, Fourier expansions of many types of clauses with great interest can be computed easily. Details can be found in the proof of Proposition \ref{prop:fourier_coefficients}. 
	
	\begin{proposition}
	Fourier expansions of \texttt{CNF}, \texttt{XOR}, \texttt{NAE} clauses and cardinality constraints have closed form representations.
	\label{prop:fourier_coefficients}
	\end{proposition}
	For a  formula $f$ with clause set $C(f)$,  we define the \emph{objective function} associated with $f$, denoted by $F_f$, by the sum of Fourier expansions of $f$'s clauses, i.e., 
	$$F_f=\sum\limits_{c\in C(f)}\texttt{FE}_c$$ 
	where $\texttt{FE}_c$ denotes the Fourier expansion of clause $c$.
	
	The degree (maximum number of variables in all the terms) of $F_f$ equals to the maximum number of literals among all clauses. 
	Note that, in general, $F_f$ is not the Fourier expansion of $f$. 
	Instead, it can be understood as a substitute of $f$'s Fourier expansion which is relatively easy to compute. 
	
	Let us provide an example to  illustrate the above ideas.
	\begin{eg}
	    Suppose $f = \left(x_1\vee x_2 \right) \wedge(x_3) \wedge\left(x_2\vee\neg x_4\right)$. 
	    Then, $C(f)=\{x_1\vee x_2,\;x_3,\;x_2\vee \neg x_4\}$ and
	    \begin{equation}\nonumber
	        \begin{split}
	       F_f&=\left(-\frac{1}{2}+\frac{1}{2}x_1+\frac{1}{2}x_2+\frac{1}{2}x_1x_2\right)+ x_3   
	        +\left(-\frac{1}{2}+\frac{1}{2}x_2-\frac{1}{2}x_4-\frac{1}{2}x_2x_4\right) \\
	          &= -1+\tfrac{1}{2}x_1+x_2+x_3-\tfrac{1}{2}x_4+\tfrac{1}{2}x_1x_2-\tfrac{1}{2}x_2x_4.
	        \end{split}
	    \end{equation}
	\end{eg}
	
	Notice that the objective function is nothing special but a polynomial.
	Therefore we relax the domain from discrete to continuous. An assignment is now defined as a real vector $a\in[-1,1]^n$. 
	The reduction is formalized in Theorem \ref{red}.
	
	\begin{theorem}{(Reduction)}
		\label{red} $f$ is satisfiable if and only if $$\mathop{\min}\limits_{x\in [-1,1]^n}F_f(x)=-m.$$
	\end{theorem}
	Recall that $m$ is  the number of clauses. Theorem \ref{red} reduces SAT to a multivariate minimization problem over $[-1,1]^n$. In the rest of this subsection, we will provide proof sketch  of Theorem \ref{red}.
 
 	\begin{definition} (Constant). A clause is constant if it is  equivalent to either {\rm \texttt{True}} or {\rm \texttt{False}}. 
	The Fourier expansion of a clause is constant if it always equals to $-1$ or $1$. 
	\end{definition}
 
	Lemma \ref{prange} indicates the value of multilinear polynomials is well-behaved in the cube $[-1,1]^n$.
     \begin{lemma} \label{prange}Let $c$ be  a non-constant clause and $a$ be an assignment. 
		Then:
		\begin{enumerate}
		\item if $a_i\in \{-1,1\} $ for all $i\in [n]$, then $\texttt{FE}_c(a)\in \{-1,1\}$.
			\item if $a_i\in [-1,1] $ for all $i\in [n]$, then $\texttt{FE}_c(a)\in [-1,1]$.
			\item if $a_i\in (-1,1)$ for all $ i\in [n]$, then $\texttt{FE}_c(a)\in (-1,1)$.
			\end{enumerate}
	\end{lemma}
	In order to formalize the procedure of converting a fractional solution to a Boolean assignment, we define the concept ``feasibility''.
	
		\begin{definition}(Partially assigned function) Suppose $[n]$ is partitioned into two sets, $J$ and $[n]-J$. Let $z\in [-1,1]^{|J|}$ be a real vector, we write $F_{J\gets z}$ for the partially assigned function of $F$ given by fixing the coordinates in $J$ to be the values $z$ in $F$.
	\end{definition}

		\begin{definition}
	\label{feasible}(Feasible Solution). For an objective function $F$ of a Boolean formula and an 
		assignment $a\in[-1,1]^n$, 
		let $I = \{i\;|\;a_i\in\{\pm 1\}\}$. $a$ is a \textbf{feasible solution} of $F$ if  $F_{I\gets a_{I}}$ is constant, where $a_{I}$ is the projected vector of $a$ at coordinates in $I$. We  say $a$ is \emph{feasible} if $F$ is clear in context.
	\end{definition}
	
	\begin{eg}
		Let $f = x_1\wedge x_2$; then 
		$$F_f= \tfrac{1}{2}+\tfrac{1}{2}x_1+\tfrac{1}{2}x_2-\tfrac{1}{2}x_1x_2.$$ $a=(1,-0.3)$ is a feasible solution because $I=\{1\}$ and $F_{I\gets a_{I}}=F_{1\gets 1}=\frac{1}{2}+\frac{1}{2}+\frac{1}{2}x_2-\frac{1}{2}x_2=1$ is constant not depending on $x_2$.
	\end{eg}
	
	By Definition \ref{feasible} if we find a feasible solution $a^\star$ of the objective function, we can adjust it into a Boolean assignment $b^\star$ by modifying values of $a^\star$ in $(-1,1)$ to $\{\pm 1\}$ in $b^\star$ arbitrarily. 
	By feasibility of $a^\star$, $F_f(a^\star)=F_f(b^\star)$.
	
		\begin{lemma} \label{coro:feasible_solution} Let $c$ be  a non-constant clause and $a$ be an assignment. 
	If $\texttt{FE}_c(a)=-1$, then $a$ is a feasible solution of $\texttt{FE}_c$.
	\end{lemma}
	
	\begin{proof*}
 Suppose $\texttt{FE}_c(a)=-1$. 
	We partition $[n]$ into two sets, $I$ and $[n]-I$, where $I = \{i\;|\;a_i\in\{\pm 1\}\}$ and $[n]-I = \{i\;|\;a_i\in(-1,1)\}$.
		
		Consider the polynomial $\texttt{FE}_{c\{I\gets a_I\}}$,
		\begin{itemize}[leftmargin=0.5cm]
			\item[--] If $\texttt{FE}_{c\{I\gets a_I\}}$ is constant, then $a$ is feasible by definition.
			\item[--] Otherwise, $\texttt{FE}_{c\{I\gets a_I\}}$ is non-constant thus $[n]-I\neq \emptyset$. Since every variable with index in $I$ is fixed in $\{\pm 1\}$, $\texttt{FE}_{c\{I\gets a_I\}}$ is a Fourier expansion of a Boolean function on $[n]-I$.
			Since $a_i\in (-1,1)$ for every $i\in [n]-I$, by the third argument of Lemma \ref{prange} we have 
			$$\texttt{FE}_c(a)=\texttt{FE}_{c\{I\gets a_I\}}(a_{[n]-I})\in (-1,1),$$ 
			which conflicts with $\texttt{FE}_c(a)=-1$. Thus $[n]-I=\emptyset$ and $a$ is feasible by definition.
		\end{itemize}
		\hfill\qedsymbol
	\end{proof*}
Now we are ready to prove Theorem \ref{red}.
\begin{proof2*}
Note that by the second argument in Lemma \ref{prange}, we have $F_f(a)\ge -m$ for all $a\in[-1,1]^n$. 
		\begin{itemize}
			\item[--] "$\Rightarrow$":  Suppose $f$ is satisfiable and $\phi\in\{\pm 1\}^n$ is one of its solution. $\phi$ is also a solution of every clause of $f$. Thus for every $c\in C(f)$, $\texttt{FE}_c(\phi)=-1$. Therefore $F_f(\phi)=-m$ and $\mathop{\min}\limits_{x\in [-1,1]^n}F_f(x)=-m$. 
			\item[--] "$\Leftarrow$": Suppose $\mathop{\min}\limits_{x\in [-1,1]^n}F_f(x)=-m$. Thus $\exists\, a^\star$ such that $F_f(a^\star)=-m$. By the second argument in Lemma \ref{prange}, we have $\texttt{FE}_c(a^\star)=-1$ for every $c\in C(f)$. By Lemma \ref{coro:feasible_solution} 
			$a^\star$ is a feasible solution of $FE_c$ for every $c\in C(f)$. Thus $a^\star$ is also a feasible solution of $F_f$. Therefore, by feasibility, we can get a Boolean assignment which satisfies all the clauses by arbitrarily rounding all the fractional coordinates. \hfill\qedsymbol
		\end{itemize}
\end{proof2*}
	
\subsection{Local Landscape of Multilinear Polynomials}
In this subsection we characterize the local landscape of multilinear polynomials. 
We also show that, in our case, local minima are meaningful and useful.

		First, we formally define local minimum for constrained problems in Definition \ref{defi:local_minimum}.
	\begin{definition} (Local minimum for constrained problem)
			\label{defi:local_minimum}
			For a vector $a\in \mathbb{R}^n$, we define $\mathcal{N}_{\delta}(a)$, the neighborhood of $a$ with radius $\delta$ as $\mathcal{N}_{\delta}(a)=\{x\;|\;\norm{a-x}_2^2\le \delta\}$. 
	Given a set $\Delta$, we say an assignment $a$ is a \textbf{local minimum of $F$ in $\Delta$} if 
	\begin{align*}
	    \exists \delta>0, \; \forall a'\in  \mathcal{N}_{\delta}(a)\cap \Delta, \;  F(a)\le F(a').
	\end{align*}
	\end{definition}

For a multivariate function, a saddle point is a point at which all the first derivatives are zero but there exist at least one positive and one negative direction. Therefore, a saddle point is not a local minimum in an unconstrained problem.
		\begin{lemma}\label{lemma:saddle} Every critical point (where all the first derivatives are zero) of a non-constant multilinear polynomial is a saddle point.
	\end{lemma}
		
		Lemma \ref{lemma:saddle} indicates for unconstrained, non-constant multilinear polynomials, no local minimum exists. 
		On the other hand, after adding the bound $[-1,1]^n$, intuitively, a local minimum of $F$ in $[-1,1]^n$ can only appear on the boundary. Lemma \ref{lemma:local} uses feasibility again to formalize this intuition.
	\begin{lemma}\label{lemma:local}  If $a^\star\in[-1,1]^n$ is a local minimum of $F$ in $[-1,1]^n$, then $a^\star$ is feasible.
	\end{lemma}

		Lemma \ref{lemma:local} indicates that every local minimum  $a^\star$ in $[-1,1]^n$ is meaningful in  the sense that it can be easily converted into a Boolean assignment $b^\star$. 
		Moreover, since $b^\star$ is a Boolean assignment, it is easy to see that the number of clauses satisfied by $b^\star$ is $\frac{m-F(a^\star)}{2}$, which is a special case of Theorem \ref{theo:rounding} in the next section. Thus if a global minimum is too hard to get, our method is still useful for some problems, e.g., MaxSAT. 
	
		In most multilinear polynomials generated from Boolean formulas in practice, saddle points rarely exist. 
		However, there exist ``pathological'' polynomials that have unaccountably infinite saddle points, as Example \ref{exam:uncountable-saddle} shows.
			\begin{eg}
	\label{exam:uncountable-saddle}
	Let $F(x)=x_1x_2x_3x_4$. Then every point which has form $(x_1,0,0,0)$, $(0,x_2,0,0)$, $(0,0,x_3,0)$ or $(0,0,0,x_4)$ is a degenerate saddle point where both the gradient and Hessian are zero.
	\end{eg}

\section{Why Continuous?}
	Before introducing our algorithm for minimizing multilinear polynomials, we would like to illustrate why doing continuous optimization might be better than discrete local search. 
	
	Our explanation is, compared to local search SAT solvers which only make progress when the number of satisfied clauses increases, our tool makes progress as long as the value of the polynomial decreases, even by a small amount.

Theorem \ref{theo:rounding} formalizes the intuition. 
It indicates that when we decrease the polynomial, we are in  fact increasing the expectation of the number of satisfied clauses, after a randomized rounding. 
	
	\begin{definition} (Randomized Rounding)
	The randomized rounding function, denoted by $R:[-1,1]^n\to \{\pm 1\}^n$ is defined by:  
	\begin{equation}\nonumber
	    \begin{cases}
	    \mathbb{P}(R(a)_i=-1)=-\frac{1}{2}a_i+\frac{1}{2}\\
	    \mathbb{P}(R(a)_i=+1)=+\frac{1}{2}a_i+\frac{1}{2}
	    \end{cases}
	\end{equation}
	\end{definition}
	where $i\in[n]$ and $a\in[-1,1]^n$. Note that the closer $a_i$ is to $-1$, the more likely $R(a)_i$ will be $-1$ and vise versa.  
	
	\begin{theorem}
	\label{theo:rounding}
	Let $f$ be a formula with $m$ clauses and $F$ be the multilinear polynomial associated with $f$. For $a\in[-1,1]^n$, let $R(a)\in \{\pm 1\}^n$ be the vector given by rounding $a$. Let $m_{\texttt{SAT}}(R(a))$ be the number of satisfied clauses by $R(a)$, then
	$$
	\mathop{\mathbb{E}}\limits_{R(a)}(m_{\texttt{SAT}}(R(a)) = \tfrac{m-F(a)}{2},
	$$
	\end{theorem}
	where $m$ is the number of clauses. In particular, if $F(a)=-m$ then $\mathop{\mathbb{E}}\limits_{R(a)}(m_{\texttt{SAT}}(R(a))=m$. Since $m_{\texttt{SAT}}(b)\le m$ for any $b\in\{\pm 1\}^n$, $m_{\texttt{SAT}}(R(a))=m$ and $R(a)$ is a solution of $f$.

	\begin{proof*}
        We first prove that for the Fourier Expansion $p$ of a clause $c$, the probability of $c$ is satisfied by $R(a)$, i.e., $\mathop{\mathbb{P}}[p(R(a))=-1]$ is $\frac{1}{2}-\frac{p(a)}{2}$. Then by the linearity of expectation, Theorem \ref{theo:rounding} follows directly.
        
        We  prove  by induction on the number of variables $n$.
	\end{proof*}
		\noindent {\rm \texttt{Basis step}}: Let $n=1$. $p$ is either constant or $x_1$, or $-x_1$. It's easy to verify the statement holds.
		
		\noindent {\rm \texttt{Inductive step}}: Suppose $n\ge 2$. Then, $p(R(a)) $ can be expanded as :
		
		\begin{align}\nonumber
		    p(R(a))&=\tfrac{1-R(a)_n}{2} \cdot p_{n \gets (-1)}(R(a)_{[n-1]})+\tfrac{1+R(a)_{n}}{2} \cdot p_{{n}\gets 1}(R(a)_{[n-1]}) \nonumber
		\end{align}
		
		Note that the value of $R(a)_n$ and $R(a)_{[n-1]}$ are independent, thus
		\begin{equation}\nonumber
		\begin{split}
		    &\mathop{\mathbb{P}}[p(R(a))=-1]\\=&\mathop{\mathbb{P}}[R(a)_n=-1]\cdot \mathop{\mathbb{P}}[p_{n \gets (-1)}(R(a)_{[n-1]})=-1] + \mathbb{P}[R(a)_n=1]\cdot \mathop{\mathbb{P}}[p_{n \gets 1}(R(a)_{[n-1]})=-1]\\
		    =&\frac{1-a_n}{2}\cdot \frac{1-p_{n\gets -1}(a_{[n-1]})}{2} + \frac{1+a_n}{2}\cdot \frac{1-p_{n\gets 1}(a_{[n-1]})}{2} \text{ (by I.H.)} \\
		    =&\frac{1}{2}-\frac{p(a)}{2}
		\end{split} 
			\end{equation}
		\hfill\qedsymbol

Research showed that local search SAT solvers, such as GSAT, spend most of the time on the so-called ``sideway'' moves \cite{GSAT}. Sideway moves are moves that do not increase or decrease the total number of unsatisfied clauses. Although heuristics and adding noise for sideway moves lead the design of efficient solvers, e.g., WalkSAT, local search SAT solvers fail to provide any guarantee when making sideway moves. 

It is illustrative to think of a cardinality constraint, e.g., the majority constraint which requires at least half of all the variables to be \texttt{True}. If we start from the assignment of all \texttt{False}'s, local search solvers need to flip at least half of all bits to make progress. In other words, local search solvers will encounter a neighbourhood with exponential size where their movements will be ``blind''. 
In contrast, guided by the gradient, our method will behave like ``flipping'' all the variables by a small amount towards the solution. 

\section{A Gradient Descent-Based Algorithm for Minimizing Multilinear Polynomials}
\label{sec:algorithm}
In this section, we propose an algorithm for minimizing multilinear polynomials associated with Boolean formulas. 
We aim to show  promising theoretical guarantees that gradient-based algorithms can enjoy. 
In practice, there are elaborate packages available and we used them in our experiments.
		
		Since the objective function is constrained, continuous and differentiable, projected  gradient descent (PGD) \cite{Nesterov:2014:ILC:2670022}  is a candidate for solving our optimization problem. As a non-convex problem, the initialization plays a key role on the result. 
	We use random initialization for each iteration and return the best result within a time limit.

		\paragraph{Efficient Evaluation of Objective Function.}
	Although theoretically the objective function can have up to $2^n$ terms, we do not actually store all the coefficients and do a naive evaluation to get $F(a)$ for $a\in[-1,1]^n$. 
	Instead, we leverage the symmetry of clauses to compute the value of Fourier expansion of each clause $c$, denoted as $\texttt{FE}_c(a)$, separately.  	By this trick, we are able to  evaluate the objective function in $O(\sum\limits_{c\in C(f)} k_c^2)$ time (in worst case $O(n^2m)$), where $k_c$ is the length of clause $c$.
	By this trick, we bypass considering data structures that store the multilinear polynomials.
	
	First note that for every negative literal, say $\neg x_i$, we can convert it to positive by flipping the sign of $a_i$, since applying a negation on formulas is equivalent to adding a minus sign to Fourier expansions (assume each variable appears at most once in a clause $c$). 
	
	Now suppose $c$ is a clause from $\{$\texttt{CNF}, \texttt{CARD}, \texttt{XOR}, \texttt{NAE}$\}$ with no negative literals. Then $c$ is symmetric, which means its Fourier coefficient at $S$ only depends on $|S|$. Thus the set of Fourier coefficients can be  denoted as $\texttt{Coef}(\texttt{FE}_c) = \left(\kappa(\emptyset),\kappa([1]),\kappa([2]),\dots,\kappa([k_c])\right)$.	Let $s(a) = \left(1,\sum_{i=1}^{k_c}a_i,\sum_{i< j}a_ia_j,\dots,\prod_{i=1}^{k_c} a_i\right)$. One can easily verify that  $\texttt{FE}_c(a) = \texttt{Coef}(\texttt{FE}_c)\cdot s(a)$, where "$\cdot$" represents the inner product of two vectors. 
	
	$s(a)$ can be obtained by computing the coefficients of $t^0$, $t^1,\cdots, t^{k_c}$ in the expansion of the following polynomial.
	
	\begin{equation}\nonumber
	    \begin{split}
        	\prod_{i=1}^{k_c} \left(a_i+t \right) =t^{k_c} + \left(\sum_{i=1}^{k_c}a_i \right) \cdot t^{{k_c}-1}  + \cdots + \prod_{i=1}^{k_c}a_i\cdot t^0	        
	    \end{split}
	\end{equation}

		The coefficients of the polynomial above can be computed in $O({k_c}^2)$ time, given $a\in[-1,1]^{k_c}$. Thus we can evaluate the objective function in $O(\sum\limits_{c\in C(f)} k_c^2)$.

Since the  gradient and Hessian of a multilinear polynomial are still multilinear \cite{O'Donnell:2014:ABF:2683783}, we are able to calculate them analytically by the method above. 
In experiments, we observed feeding gradients to the optimizer significantly accelerated the minimization process.
	
	\paragraph{Our PGD-Based Multilinear Optimization Algorithm.}
	In Algorithm \ref{algo:main_algorithm},  we propose a PGD-based algorithm for multilinear optimization problem. 
	In   gradient descent, the iteration for minimizing an objective function $F$ is:
	$$
	x_{t+1}' = x_{t}-\eta \cdot \nabla F(x_t), \;\; x_{t+1} = x_{t+1}',
	$$
	where $\eta>0$ is a step size.
	
	For a constrained domain other than $\mathbb{R}^n$, $x_{t+1}'$ may be outside of the domain. 
	In PGD, we choose the point nearest to $x_{t+1}'$ in $[-1,1]^n$ as $x_{t+1}$ \cite{NiaoHeLectureNotes}, i.e., the Euclidean projection of $x_{t+1}'$ onto the set $[-1, 1]^n$, denoted as $\Pi_{[-1,1]^n}(x_{t+1}')$.
	\begin{definition}
	The Euclidean projection of a point $y$, onto a set $\Delta$, denoted by $\Pi_{\Delta}(y)$, is defined as
	$$
	\Pi_{\Delta}(y) = \argmin_{x\in \Delta} \tfrac{1}{2} \|x-y\|^2_2.
	$$
	\end{definition}
	In our case $\Delta \equiv [-1,1]^n$; computing such a projection is almost free, as shown in Proposition \ref{projection}.
	\begin{proposition}
	\label{projection}
	$$\Pi_{[-1,1]^n}(y)_i=
	\begin{cases}
	y_i, & \text{if } y_i \in [-1,1], \\
	{\rm \rm{\texttt{sgn}}}(y_i), & \text{otherwise}.
	\end{cases}
	$$
	\end{proposition}
	
	Combining the above, the main iteration of PGD can be rewritten as
	$$
	x_{t+1} = \Pi_{[-1,1]^n}\left(x_t-\eta \cdot \nabla F(x_t)\right) = x_t-\eta G(x_t),$$
	where $G(x)=\tfrac{1}{\eta}\left(x_t-\Pi_{[-1,1] ^n}\left(x_t-\eta \nabla F(x_t)\right)\right)$ is regarded as the \textit{gradient mapping}.

	\begin{algorithm}[t!]
    \SetAlgoLined
    \SetKwInOut{Input}{Input}
    \SetKwInOut{Output}{Output}
    \Input{Polynomial $F$, $\eta > 0$, $\varepsilon > 0$.}
    \Output{Approximately minimizer $\widehat{x}$.}
    \vspace{0.1cm}
    \hrule
    \vspace{0.1cm}
    \For{$j=1,\; \dots, \;J$}{
        $x_0 \sim \mathcal{U}[-1,1]^n$\\
        \For{ $t = 0,\; \dots $}
        {
            $G(x_t)=\frac{1}{\eta}\left(x_t -\Pi_{[-1,1]^n}\left(x_t-\eta \nabla F(x_t)\right)\right)$\\
            \eIf{$\|G(x_t)\|_2 > 0$}
            {
                $x_{t+1} = x_t - \eta\cdot  G(x_t)$\\
            }
            {
                \eIf{$x_t$ \texttt{not feasible} }
                {
                    $x_{t+1} =  \texttt{DecInnerSaddle}(F,\;x_t, \;\eta)$
                } 
                {
                    $I = \{i \; | \;(x_t)_i\in\{-1,1\}\}$\\
                    \eIf{ $\nabla F(x_t)_i\neq0, \; \forall i\in I$}
                    {
                        \texttt{Break} \hfill // \texttt{Prop. \ref{prop:localmin-firstorder}} \label{line:firstorderlocalmin}
                    }
                    {
                        $(\texttt{LocalMinFlag}, \;x_{t+1}) =  \texttt{useHessian}(F, x_t)$\\ 
                        \If{$\texttt{LocalMinFlag} = $\texttt{True} or \texttt{Unknown}}
                        {
                            \texttt{Break}
                        }
                    }
                }
            }
        }
        \textbf{Until convergence}
    }
    \Return  $x_j$ with the lowest $F(x)$ after $J$ iterators
 \caption{PGD for multilinear $F$}
 \label{algo:main_algorithm}
\end{algorithm}

In Algorithm \ref{algo:main_algorithm}, we start at a uniformly random point in $[-1,1]^n$. 
When gradient mapping $G(\cdot)$ is large, we follow it to decrease the function value. 
Otherwise, it means the algorithm either reaches a local minimum in $[-1,1]^n$, or falls into a saddle point where the original gradient $\nabla F(\cdot)$ is  negligible. 
If the first case happens we are done for this iterator. 
Else, we still try to escape the saddle point by additional methods.

In Theorem \ref{theo:convergence_rate}, we show that Algorithm \ref{algo:main_algorithm} is guaranteed to converge to a point where the projected mapping is small, $\|G(x_t)\|_2 = 0$,\footnote{In practice, a stopping criterion to use is $\|G(x_t)\|_2 < \varepsilon$ for a small accuracy level $\varepsilon > 0$.} and depends polynomially on $n$, $m$ and tolerance $\varepsilon$. In the proof, we used techniques from \cite{Stochastic_Gradient_Descent_Escapes_Saddle_Points_Efficiently} and \cite{NiaoHeLectureNotes}.
	
	\begin{theorem} (Convergence speed)
	\label{theo:convergence_rate}With the step size $\eta=\frac{1}{nm}$, Algorithm \ref{algo:main_algorithm} converges to a $\epsilon$-projected-critical point (where $\|G(x)\|_2<\epsilon$) in  $O(\frac{nm^2}{\epsilon^2})$ iterators in the worst case.
	\end{theorem}
The proof of Theorem \ref{theo:convergence_rate} relies on the fact that multilinear polynomials on $[-1,1]^n$ are relatively smooth as Proposition \ref{lipschitz} indicates.
\begin{proposition} (Lipschitz)
	\label{lipschitz}
	Let $F:[-1,1]^n\to [-m,m]$ be a multilinear polynomial. Then, for every $x,y\in [-1,1]^n$,
	\begin{enumerate}
	    \item $|F(x)-F(y)|\le m \sqrt{n} \cdot \|x-y\|_2$. I.e., $F$ is $(m \sqrt{n})$-Lipschitz continuous.
	    \item $ \|\nabla F(x)-\nabla F(y)\|_2\le m n \cdot \|x-y\|_2$.  I.e., $F$ has $(m n)$-Lipschitz continuous gradients.
	    \item $\|\nabla^2 F(x)-\nabla^2 F(y)\|_2 \le \left(m n^{\frac{3}{2}}\right) \cdot \|x-y\|_2$. I.e., $F$ has $\left(m n^\frac{3}{2}\right)$-Lipschitz Hessians.
	\end{enumerate}
		All the bounds are tight if we consider the parity function $F(x)=\Pi_{i\in[n]}x_i$.
		\end{proposition}

Proposition \ref{prop:localmin-firstorder} shows that in Algorithm \ref{algo:main_algorithm} we can identify local minima by gradient information. 
	\begin{proposition}
	\label{prop:localmin-firstorder}
	When algorithm \ref{algo:main_algorithm} reaches line \ref{line:firstorderlocalmin}, $x$ is a local minimum. i.e, if $x$ is feasible, $G(x)=0$ and $\nabla F(x)_i\neq0$ for all $i\in I$, then $x$ is a local minimum in $[-1,1]^n$ of $F$.
	\end{proposition}
	We design \texttt{DecInnerSaddle} and \texttt{useHessian} to escape saddle points. \texttt{DecInnerSaddle} uses the idea from Lemma \ref{lemma:saddle} to give a negative direction, as long as the point is not feasible. 
	\texttt{useHessian} leverages second order derivatives to give a negative direction, or identifies a local minimum when the first order derivatives are too small. Due to space limit, we leave subprocedures \texttt{DecInnerSaddle} and \texttt{useHessian}, as well as their properties, in the Appendix.
	
		\subsection{Weighted Case}
		\label{subsection:weight_case}
	
	The weighted version of the objective function  is defined as
    $$
	F_f=\sum_{c\in C(f)}w_f(c)\cdot \texttt{FE}_c
	$$
	where $w_f:C(f)\to \mathbb{R}$ is the weight function. 
	It is easy to verify that the weighted version of Theorem \ref{red} and Lemma \ref{lemma:local} still hold. The weighted case is useful in two cases: 
	\begin{itemize}
	    \item \textbf{Vanishing Gradient.} When a clause contains too many of variables (e.g., a global cardinality constraint), we observed that the gradient given by this clause on a single variable becomes negligible at most points. By assigning large weights to long clauses, gradient varnishing can be alleviated. 
	    \item \textbf{Weighted MAX-SAT.} By returning a local minimum, our tool can be used to solve weighted MAX-SAT problems.
	\end{itemize}

\section{Experimental Results}
In this section we compare our tool, \texttt{FourierSAT} with other SAT solvers on random and realistic benchmarks. The objective of our experimental evaluation is to
answer these three research questions:
\\
\textbf{RQ1.} How does \texttt{FourierSAT} compare to local search SAT solvers with cardinality constraints and \texttt{XOR}s encoded in \texttt{CNF}?  \\
\textbf{RQ2.} How does \texttt{FourierSAT} compare to specialized \texttt{CARD}-\texttt{CNF} solvers with respect to handling cardinality constraints?  \\
\textbf{RQ3.} How does \texttt{FourierSAT} compare to CDCL SAT solvers with cardinality constraints/\texttt{XOR}s encoded in \texttt{CNF}?

\subsection{Experimental Setup}
We choose \texttt{SLSQP} \cite{SLSQP}, implemented in  \texttt{scipy} \cite{scipy} as our optimization oracle, after comparing available algorithms for non-convex, constrained optimization.

 Since random restart is used, different iterators are independent. 
 Thus, we parallelized \texttt{FourierSAT} to take advantage of multicore computation resources. 
 Each experiment was run on an exclusive node in a homogeneous Linux cluster. These nodes contain 24-processor cores at 2.63 GHz each with 1 GB of RAM per node. 
 The time cap for each job is 1 minute.

We compare our method with the following SAT solvers:
\begin{itemize}
    \item Cryptominisat \cite{cmspaper}  (CMS), an advanced SAT solver supporting \texttt{\texttt{CNF}}+\texttt{XOR} clauses by enabling Gauss Elimination.
      \item WalkSAT \cite{walksat}, an efficient local search SAT solver. We used a fast C implementation due to \cite{Walksat-implementation}.
    \item MiniCARD \cite{minicard}, a MiniSAT-based \texttt{CARD}-\texttt{CNF} solver. It has been implemented in  \texttt{pysat} \cite{pysat}. 
    \item MonoSAT \cite{monosat2015}, a SAT Modulo Theory solver which supports a large set of graph predicates.
\end{itemize}
MiniCARD can handle cardinality constraints natively. For MonoSAT, we reduced cardinality constraints to max-flow. 
For CMS and WalkSAT, we applied a set of cardinality encodings, which includes Sequential Counter \cite{sequential-counter}, Sorting Network \cite{sort-network}, Totalizer \cite{totalizer} and Adder \cite{adder}. 
For solvers that do not support \texttt{XOR}s, we used a linear encoding due to \cite{xor-encoding} to decompose them into 3-\texttt{CNF}.
To address vanishing gradient, the weight of each clause equals to its length. 

\subsection{Benchmarks}
    We generated hybrid Boolean constraints using natural encodings of the following benchmark problems.
	\paragraph{Benchmark 1: Approximate minimum vertex covering.} Minimum vertex covering is a well-known NP-optimization problem (NPO) problem.  For each $n\in \{50,100,150,200,250\}$, we generated 100 random cubic graphs. For each graph, \texttt{Gurobi} \cite{gurobi} was used to compute the minimum vertex cover, denoted by $\texttt{Opt}$. Then we added one cardinality constraint, $D^{\le k}(X)$ where $k= \ceil{1.1\cdot|\texttt{Opt}|}$ to the \texttt{CNF} encoding of vertex covering.
	
	\paragraph{Benchmark 2: Parity learning with error.} Parity Learning 
	is to identify an unknown parity function given I/O samples. Technically, in this task one needs to find a solution of an XOR system while tolerating some constraints to be violated. Suppose there are $n$ candidate variables and $m$ I/O samples (XOR constraints). A solution to this problem is allowed to be incorrect on at most $(e\cdot m)$ I/O samples. For $e=0$ the problem is in P (Gaussian Elimination); for $0<e<\frac{1}{2}$, whether the problem is in P still remains open. Parity learning is a well-known hard for SAT solvers as well, especially for local search solvers \cite{parity-learning-is-hard}. 
	
We chose $N\in\{8,16,32\}$, $e=\frac{1}{4}$ and $m=2N$ to generate hard cases. For \texttt{FourierSAT}, this problem can be encoded into solving $M$ \texttt{XOR} clauses where we  allow at most $eM$ clauses to be violated. For WalkSAT and CMS, we used the encoding due to \cite{satlib}. 
	
	\paragraph{Benchmark 3: Random \texttt{CNF}-\texttt{XOR}-\texttt{CARD} formulas.}  For each $n\in\{50,100,150\}$,  $r\in\{1,1.5\}$, $s\in \{0.2,0.3\}$, $l\in \{0.1,0.2\}$ and $k\in\{0.4n,0.5n\}$, we generated random benchmarks with $rn$ 3-\texttt{CNF} clauses, $sn$ \texttt{XOR} clauses with length $ln$ and 1 global cardinality constraint $D^{\le k}(X)$. Those parameters are chosen to guarantee that all problems are satisfiable \cite{cnfxor-phase-transition,cnfxor-card-phase-transition}.

\begin{figure}[t]
\centering
\includegraphics[width=0.65\columnwidth]{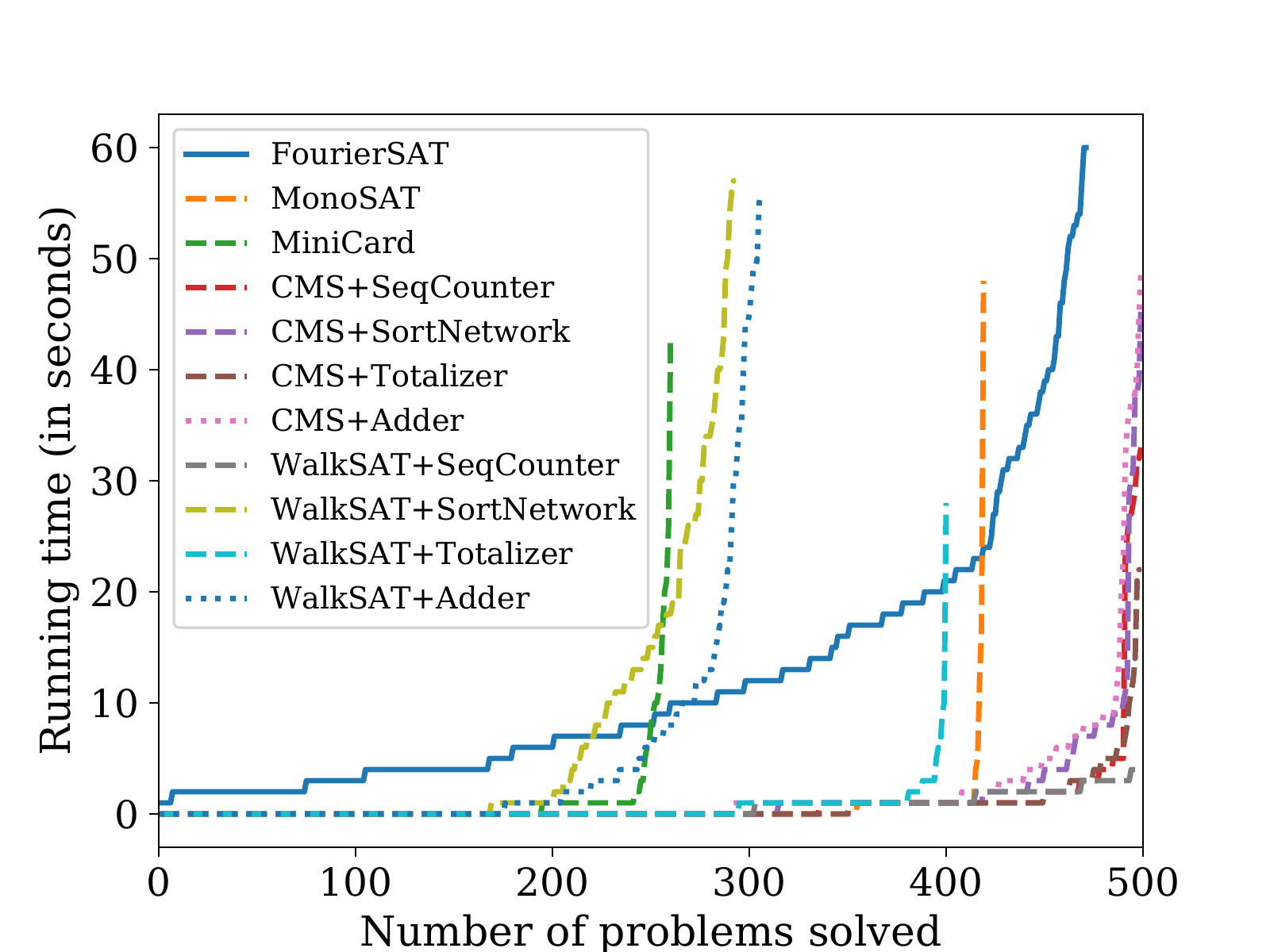}
\caption{Results on 500 vertex covering problems}
\label{exp:vc}
\end{figure}

\begin{figure}[t]
\centering
\includegraphics[width=0.65\columnwidth]{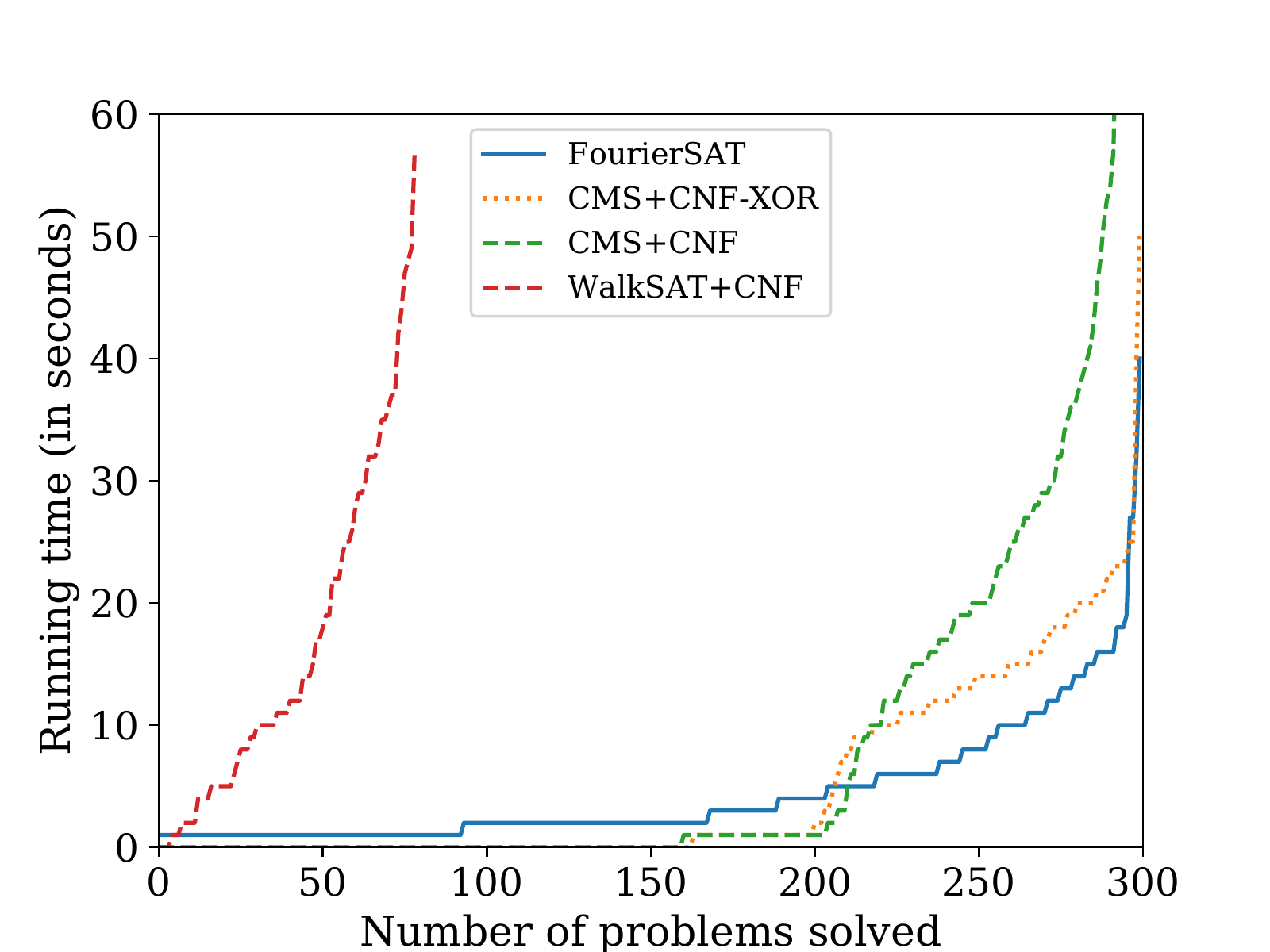}
\caption{Results on 300 parity learning with error problems}
\label{exp:par}
\end{figure}

\begin{figure}[t]
\centering
\includegraphics[width=0.65\columnwidth]{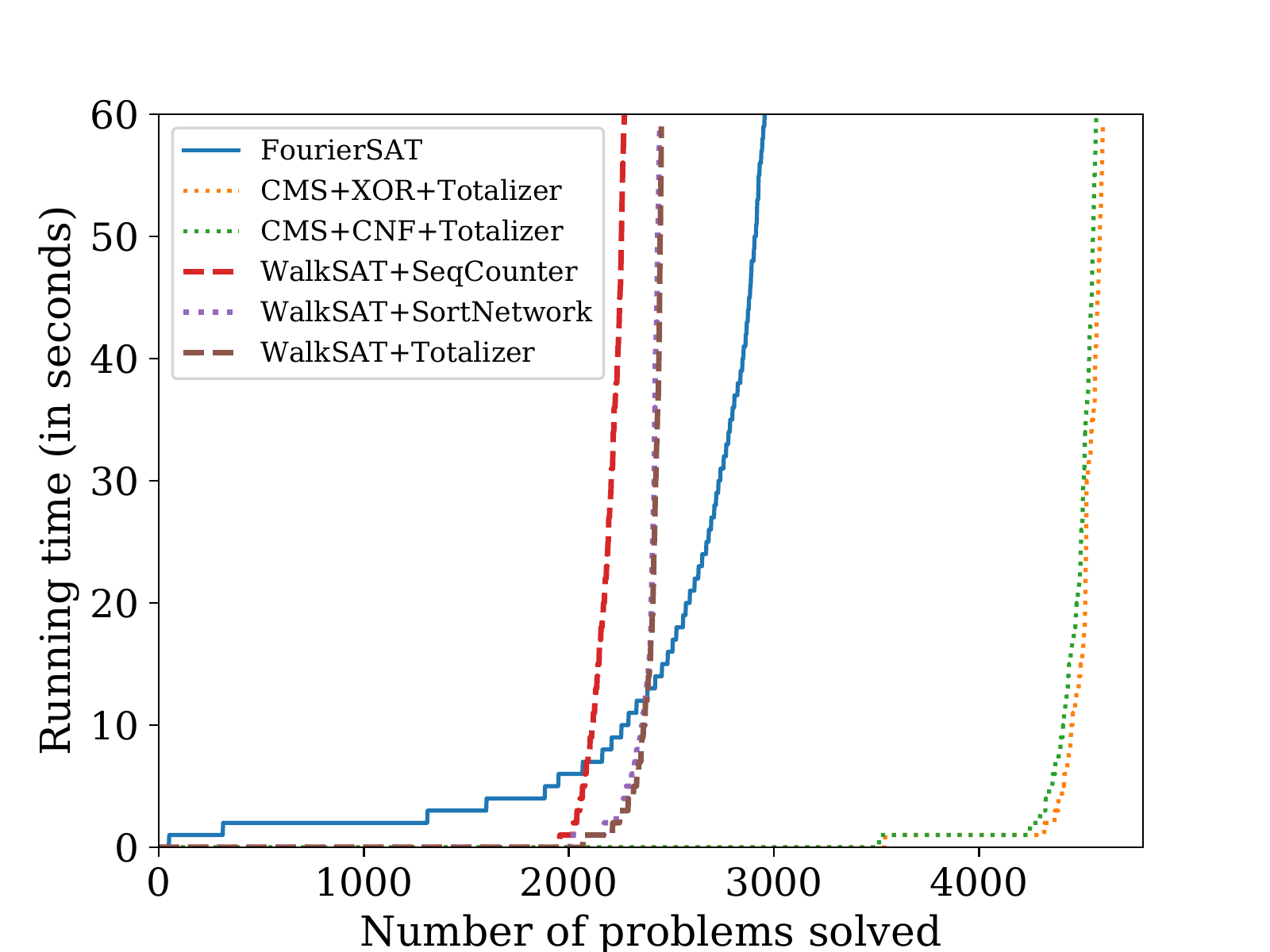}
\caption{Results on 4800 random \texttt{CNF}-\texttt{XOR}-\texttt{CARD} problems}
\label{exp:hybrid}
\end{figure}

\subsection{Results}
The experimental results of three benchmarks above are shown in Figure \ref{exp:vc}, \ref{exp:par} and \ref{exp:hybrid} respectively. 

\noindent\textbf{RQ1.} Figure \ref{exp:vc} shows \texttt{FourierSAT} solved more problems than WalkSAT did with all the cardinality encodings except SeqCounter. For parity learning problem, WalkSAT with \texttt{CNF} encoding only solved 79 problems, while \texttt{FourierSAT} was able to solved all 300 problems. For the random hybrid constraints benchmarks shown in Figure \ref{exp:hybrid}, \texttt{FourierSAT} solved 2957 problems while WalkSAT with SeqCounter, SortNetwork and Totalizer solved 2272, 2444 and 2452 problems  respectively. In our experiment, \texttt{FourierSAT} is more robust than WalkSAT, which agrees on our intuition in Section 4.
\\
\textbf{RQ2.} In Figure \ref{exp:vc} it is clear that \texttt{FourierSAT} solved more problems (472) than MonoSAT (420) and MiniCard (261) did, which indicates that \texttt{FourierSAT} is capable of handling cardinality constraints efficiently.
\\
\textbf{RQ3.} For the parity learning problem shown in Figure \ref{exp:par}, the competition of \texttt{FourierSAT} and CMS with \texttt{CNF}+\texttt{XOR} formula is roughly a tie, while \texttt{FourierSAT} outperformed CMS with pure \texttt{CNF} formula. For other two benchmarks shown in Figure \ref{exp:vc} and \ref{exp:hybrid}, we observed that CMS has the best performance, especially for solving random hybrid constraints, despite of the choice of encodings. 

Due to the maturity of CMS, it is not surprising that CMS performed better than \texttt{FourierSAT} did on some benchmarks, especially when dealing with long \texttt{XOR} clauses. However, the experimental result shows that as a versatile solver, \texttt{FourierSAT} is comparable with many existing, well-developed, specialized solvers.

Furthermore, we notice that although \texttt{FourierSAT} is usually slower than other solvers on easy problems, it seems to scale better. One potential reason is, due to the continuous nature of \texttt{FourierSAT}, it avoids suffering from scaling exponentially too early. This behavior of \texttt{FourierSAT} increases our confidence in that algebraic methods are worth exploring for SAT solving in the future.

\section{Conclusion and Future Directions}
In this paper, we propose a novel algebraic framework for solving Boolean formulas consisting of hybrid constraints. Fourier expansion is used to reduce Boolean Satisfiability to multilinear optimization. Our study on the landscape and properties of multilinear polynomials leads to the discovery of attractive features of this reduction. Furthermore, to show the algorithmic benefits of this reduction, we design algorithms with certain theoretical guarantee. Finally, we implement our method as \texttt{FourierSAT}. The experimental results indicate that in practical, \texttt{FourierSAT} is comparable with state-of-the-art solvers on certain benchmarks of hybrid constraints by leveraging parallelization.

We also list a few future directions in the following:
\begin{itemize}
    \item \textbf{Complete algebraic SAT solvers.} Besides giving solutions to satisfiable formulas, we also aim to prove unsatisfiability algebraically. In other words, we hope to design a complete SAT solver based on algebraic approaches.

    Several theoretical tools, such as Hilbert's Nullstellensatz and Gr{\"o}ebner basis (e.g., see \cite{herbert}) can be used for giving an algebraic proof of unsatisfiability. Previous algorithmic work on this direction considered specific polynomial encodings on  problems such as graph coloring \cite{herbert-thesis,herbert-davis}. We are interested to see the behavior of these algebraic tools on more problems with Fourier transform as the encoding in practice.
    
    \item \textbf{Escaping local minima and saddle points.} As a local-information-based approach, \texttt{FourierSAT} suffers from getting stuck in local minima and relies heavily on random restart as well as parallelization. We believe the idea of ``sideway moves" from Local Search SAT solvers \cite{GSAT} is a good direction to address these issues. For example, one can use the solution of WSAT/FourierSAT  as the initial point of FourierSAT/WSAT. We will also consider purely first-order gradient descent approaches, probably with random noise \cite{Stochastic_Gradient_Descent_Escapes_Saddle_Points_Efficiently}, and how they perform on escaping saddle points both in theory and practice.
    \item \textbf{Alternative objective functions.} The Fourier expansions of clauses with large number of variables and low solution density can be much less smooth, which is one of the main difficulties for FourierSAT in practice. To refine the objective function, one possible way is to develop better weight functions, which can be static or dynamic. 
    
    Due to the techniques for learning Fourier coefficients \cite{Polynomial-Threshold-Functions-AC0-Functions,Constant-depth-circuits-Fourier-transform-and-learnability}, it is also promising to use the low-degree approximation of Fourier expansions as the objective function.
\end{itemize}
  
   \section*{Acknowledgement}
 Work supported in part by NSF grants IIS-1527668, CCF-1704883, and IIS-1830549.

\bibliographystyle{unsrt}  
\bibliography{ref} 
\newpage
\section*{Appendix}
\subsection{Proof of Proposition \ref{proposition:pseudo}}
	\begin{pfs*}We reduce counting the number of solutions of a knapsack problem, a \#P-hard problem, to the computation of the constant term of the Fourier expansion of a pseudo-Boolean function.
	\end{pfs*}
We reduce the counting of the number of solutions of a knapsack problem, a \#P-hard problem, to the computation of the constant term of the Fourier expansion of a pseudo-Boolean function. 
		
	For a knapsack problem with weights of items $w_1,w_2,...,w_n$ and capacity $c$, we construct the pseudo-Boolean constraint $C$: $\sum_{i}w_ix_i\ge \sum_{i}w_i-2c$ where $x\in\{\pm 1\}^n$. It is easy to see that there is a bijection between solutions of $C$ and the knapsack problem. Let the corresponding Boolean function of $C$ be $F_C$ ($F_C=-1$ if $C$ is satisfied) and the number of solutions of $C$ be $N(C)$. By Theorem 1, the constant term $\hat{F}_C(\emptyset)=\mathop{\mathbb{E}}\limits_{x\sim\{\pm 1\}^n}[F_C]=\frac{(-1)\cdot N(C)+1\cdot(2^n-N(C))}{2^n}$ and $N(C)={2^{n-1}(1-\hat{F}_C (\emptyset))}$. Thus there is a reduction from counting the solutions of a knapsack problem to computing Fourier Coefficients of a pseudo-Boolean function.\hfill\qedsymbol

\subsection{Proof of Proposition \ref{prop:fourier_coefficients}}
\paragraph{Fourier expansion of cardinality constraints}
		$$
		\widehat{D}^{\ge k}(\emptyset) =1-\frac{\sum_{i=k}^{n}\binom{n}{i}}{2^{n-1}}
		$$
			For every nonempty $S\subseteq [n]$, let $\theta$ be an arbitrary variable, the Fourier coefficient of $D^{\ge k}(S)$ is given by:
		$$
		\widehat{D}^{\ge k}(S) = \frac{\binom{n-1}{k-1}\big((1+\theta)^{n-k}(1-\theta)^{k-1}\big)_{[\theta^{|S|-1}]}}{\binom{n-1}{|S|-1}2^{n-1}},
		$$ 
		where $\big((1+\theta)^{n-k}(1-\theta)^{k-1}\big)_{[\theta^{|S|-1}]}$ is the coefficient of $\theta^{|S|-1}$ in the expansion of polynomial $\big((1+\theta)^{n-k}(1-\theta)^{k-1}\big)$. 	

\paragraph{Derivation of Fourier expansion of cardinality constraints}
		The following derivation is a generalization of that of Theorem 5.19 in \cite{O'Donnell:2014:ABF:2683783}. 
		
    For the case $|S|=0$. 
	By the formula of computing Fourier coefficients in Theorem 1, the constant term $\widehat{D}^{\ge k}(\emptyset)$ equals to: 
	\begin{equation}\nonumber
	\begin{split}
	 \widehat{D}^{\ge k}(\emptyset) &= \mathop{\mathbb{E}}\limits_{x\sim\{\pm 1\}^n}\left[D^{\ge k}(x)\right]\\
	  &= \tfrac{1}{2^n} \left(\mathop{\#}\limits_{x\in\{\pm 1\}^n}\textrm{$x$ has no less than $i$ ($-1$)'s}\right)\cdot(-1) \\
	  	    &+\tfrac{1}{2^n}\left(\mathop{\#}\limits_{x\in\{\pm 1\}^n}\textrm{$x$ has less than $i$ ($-1$)'s})\right)\cdot 1\\
	  	        &= \frac{\sum_{i=k}^{n}\binom{n}{i}}{2^n}\cdot(-1)+\left(1-\frac{\sum_{i=k}^{n}\binom{n}{i}}{2^n}\right)\cdot 1\\
	     &= 1-\frac{\sum_{i=k}^{n}\binom{n}{i}}{2^{n-1}}
	\end{split}    
	\end{equation}

	For $|S|\neq 0$, we compute the derivative of $D^{\ge k}(x)$, denoted by $\nabla_nD^{\ge k}(x)$.
	$$
	\nabla_nD^{\ge k}(x)=\frac{1}{2}((D^{\ge k}_{n\gets 1}(x)) -(D^{\ge k}_{n\gets -1}(x)))
	$$
	Since $D^{\ge k}(x)$ is a monotonic function, $\nabla_nD^{\ge k}(x)$ is a function from $\{\pm 1\}^{n-1}$ to $\{0,1\}$, the $0$-$1$ indicator of the set of $(n-1)$-bits strings with exactly $k-1$ coordinates equal to $-1$. By the derivative formula and the fact that $D^{\ge k}(x)$ is symmetric, 
	$$
	\widehat{D^{\ge k}}(S)= \widehat{\nabla_nD^{\ge k}}(T)
	$$
	for any $T\subseteq[n-1]$ with $|T|=|S|-1$.
	
	By the probabilistic definition of $T_\rho$, we have
	
	\begin{equation}\nonumber
	\begin{split}
	&T_\rho (\nabla_nD^{\ge k})(1,1,...,1)
	=\mathop{E}\limits_{x\sim N_p(1,1,...,1)}[\nabla_nD^{\ge k}(x)]=\mathop{\mathbb{P}}[x\text{ has $(k-1)$ $-1$'s}]
		\end{split}
	\end{equation}
	where each coordinate of $x$ is 1 with prob. $\frac{1}{2}+\frac{1}{2}\rho$. Thus
	$$
	T_\rho (\nabla_nD^{\ge k})(1,1,...,1)=\binom{n-1}{k-1}(\frac{1}{2}+\frac{1}{2}\rho)^{k-1}(\frac{1}{2}-\frac{1}{2}\rho)^{n-k}
	$$
	On the other hand, by the Fourier formula for $T_\rho$ and the fact that $\nabla_nD^{\ge k}$ is symmetric we have
	\begin{equation}\nonumber
	\begin{split}
	&T_\rho (\nabla_nD^{\ge k})(1,1,...,1)=\sum_{U\subseteq [n-1]}\widehat{\nabla_nD^{\ge k}}(U)\rho^{|U|}=\sum_{i=0}^{n-1}\binom{n-1}{i}\widehat{\nabla_nD^{\ge k}}([i])\rho^i
	\end{split}
	\end{equation}
	
   The equation for computing Fourier coefficients can be obtained by equating coefficients of $\rho$.

	\paragraph{Fourier Coefficients of CNF clauses}
	   First regard all negative literals as positive ones. Since a CNF clause with all positive literals is a special case of cardinality clause, its Fourier expansion can be computed. 
	   
	   Note that the Fourier expansion of $\neg x_i$ is $-x_i$. Thus for each negative literal $\neg x_i$, we just need to flip the sign of all Fourier coefficients $\widehat{F}(S)$ where $i\in S$.
	    
	  \paragraph{Fourier Coefficients of XOR clauses}
	  The Fourier expansion of an XOR clause is simply the product of all its literals. 
	
	\paragraph{Fourier Coefficients of NAE clauses.}

	\begin{equation}\nonumber
	\widehat{\texttt{NAE}}(S)=
	    \begin{cases}
	    0, |S| \text{ is odd}\\
	    (\frac{1}{2})^{n-2}-1, |S|=0\\
	    (\frac{1}{2})^{n-2}, |S| \text{ is even and non-zero}
	    \end{cases}
	\end{equation}

\paragraph{Derivation of Fourier expansion of NAE clauses}
	Note that NAE is a even function, i.e., $\texttt{NAE}(x)=\texttt{NAE}(-x)$, thus  $\widehat{\texttt{NAE}}(S)=0$ for all $S$ s.t. $|S|$ is odd. 
	
	For $S$ with even cardinality:
	\begin{itemize}
	    \item if $S=\emptyset$, by Theorem \ref{FourierTransformation},
	    	\begin{equation}\nonumber
	\begin{split}
	    \widehat{\texttt{NAE}}(\emptyset) &= \mathop{\mathbb{E}}\limits_{x\sim\{\pm 1\}^n}\left[\texttt{NAE}(x)\right]
	    = \tfrac{1}{2^n}\left( 2+(-1)\cdot (2^n-2) \right) =(\frac{1}{2})^{n-2}-1\\
	\end{split}    
	\end{equation}
	\item else if $|S|$ is even and non-zero,
	
	 	\begin{equation}\nonumber
	\begin{split}
	    \widehat{\texttt{NAE}}(S) &= \frac{1}{2^n}\sum_{x\sim\{\pm 1\}^n}\left[\texttt{NAE}(x)\cdot \chi_S(x)\right]
	    = \tfrac{1}{2^n}\left( 2+(-1)\cdot \sum_{\substack{x\in \{\pm 1\}^n\\x\not \in \{\pm \mathbf{1}\}}}\chi_S(x) \right) \\
	\end{split}    
	\end{equation}
	 Since $\mathop{\sum}\limits_{x\in \{\pm 1\}^n}\chi_S(x)=0$ for all $S\neq \emptyset$, $\sum\limits_{\substack{x\in \{\pm 1\}^n\\x\not \in \{\pm \mathbf{1}\}}}\chi_S(x)=0-1-(-1)^{|S|}=-2$.
	  Thus, for $S$ with even cardinality,
	  $$
	  \widehat{\texttt{NAE}}(S) = (\frac{1}{2})^{n-2}
	  $$
	\end{itemize}
\hfill \qedsymbol

To prove Theorem \ref{red}, we will need to first introduce and prove Lemma \ref{prange} and Corollary \ref{coro:feasible_solution}.

Lemma 3 indicates multilinear polynomials are well-behaved in the cube $[-1,1]^n$.

\subsection{Proof of Lemma \ref{prange}}
	\begin{pfs*}
		For the sake of simplicity, let $p$ denote $FE_c$. The first argument is a direct result of the definition of the Fourier expansion of Boolean functions.
        We  proved the second and the third arguments by induction on the number of variables $n$ using the decomposition $p(a)=\tfrac{1-a_n}{2} \cdot p_{n \gets (-1)}(a_{[n-1]})+\tfrac{1+a_{n}}{2} \cdot p_{{n}\gets 1}(a_{[n-1]})$.
	\end{pfs*}
	
	The first argument is a direct result of the definition of the Fourier expansion of Boolean functions.
        We will prove other two arguments by induction on the number of variables $n$.
		
		\medskip 
		\noindent {\rm \texttt{Basis step}}: Let $n=1$. Since $p$ is non-constant and $p(1),\; p(-1)\in\{\pm 1\}$, $p$ is either $x_1$ or $-x_1$. Then, the two last arguments of the lemma hold trivially.
		
		\medskip 
		\noindent {\rm \texttt{Inductive step}}: Suppose $n\ge 2$. Then, $p(a) $ can be expanded as :
		
		$$p(a)=\tfrac{1-a_n}{2} \cdot p_{n \gets (-1)}(a_{[n-1]})+\tfrac{1+a_{n}}{2} \cdot p_{{n}\gets 1}(a_{[n-1]})$$
		\begin{itemize}[leftmargin=0.5cm]
			\item[$i)$] Suppose $a_i\in [-1,1] $ for $\forall i\in [n]$.
			\begin{itemize}	
				\item If $p_{n \gets (-1)}(a_{[n-1]})\cdot p_{n\gets 1}(a_{[n-1]})< 0$. Then,
				\begin{equation}\nonumber
				    \begin{split}
				        \left|p(a)\right| \le &\max\Big\{\tfrac{1-a_n}{2}\cdot \left|p_{n \gets (-1)}(a_{[n-1]})\right|,
				        \tfrac{1+a_n}{2}\cdot\left|p_{n\gets 1}(a_{[n-1]})\right| \Big\} \le 1
				    \end{split}
				\end{equation}
				\item If $p_{n \gets (-1)}(a_{[n-1]})\cdot p_{n\gets 1}(a_{[n-1]})\ge 0$, then we have,
			
			    	\begin{equation} \nonumber
				\begin{split}
				\left|p(a)\right| \le \left(\tfrac{1-a_n}{2}+\tfrac{1+a_n}{2}\right)\cdot &\max \Big\{\left|p_{n \gets (-1)}(a_{[n-1]})\right|,
					\left| p_{n\gets 1}(a_{[n-1]})\right|\Big\}\le 1 
				\end{split}
				\end{equation}
				Note that the last steps in the two cases follow by inductive hypothesis.
			\end{itemize} 
			\item[$ii)$] Suppose $a_i\in (-1,1) $ for $\forall i\in [n]$.
			\begin{itemize}
				\item If $p_{n \gets (-1)}(a_{[n-1]})\cdot p_{n\gets 1}(a_{[n-1]})< 0$, then, since $|\frac{1+a_n}{2}|<1,|\frac{1-a_n}{2}|<1$, what we have by inductive hypothesis is,
					\begin{equation}\nonumber
				    \begin{split}
				\left|p(a)\right| \le &\max\Big\{\tfrac{1-a_n}{2}\cdot \left|p_{n \gets (-1)}(a_{[n-1]})\right|,\tfrac{1-a_n}{2}\cdot \left|p_{n\gets 1}(a_{[n-1]})\right| \Big\}< 1.
				\end{split}
				\end{equation}
				\item If $p_{n \gets (-1)}(a_{[n-1]})\cdot p_{n\gets 1}(a_{[n-1]})\ge 0$, since we already have proved that $|p(a)|\le 1$, what we need to show is $|p(a)|\not = 1$. Note that for $|p(a)|= 1$ to be true, both $|p_{n \gets (-1)}|=1$ and $|p_{n\gets 1}|=1$ must hold. 
				By inductive hypothesis, if one of $p_{n \gets (-1)}$ and $p_{n\gets 1}$ is non-constant, then 
				$$|p_{n \gets (-1)}|<1 \text{ or } |p_{n\gets 1}|<1.$$ Therefore both $p_{n \gets (-1)}$ and $p_{n\gets 1}$ are constant and $p$ only relies on one variable, which is covered by the basis. 
			\end{itemize}
		\end{itemize}
		\hfill\qedsymbol

\subsection{Proof of Lemma \ref{lemma:saddle}}
	\begin{pfs*}
   	Intuitively, the basic idea of this proof is to show that, for every critical point, there exist two directions such that moving by a small step towards the first one will increase the function value and towards the second one will decrease the value. 
	\end{pfs*}
	In order to simplify the proof, for a multilinear polynomial $F$ and its critical point $a$, we define the following polynomial $F_{a}^0:$
		
		$$
		F_{a}^0(x)=F(x+a)-F(a).
		$$ 
		Notice that  $F_{a}^0(0) = F(a) - F(a) = 0$. It is obvious that if $0$ is a saddle point of  $F_{a}^0$ holds, then $a$ is a saddle point of $F$.

		Instead of directly proving things about arbitrary multilinear polynomials, we will prove that, for every non-constant multilinear function $f$ with $f(0)=0$, there exist $\epsilon>0$ and $v^+,v^-\in \mathbb{R}^n$, such that, for every $0<\delta<\epsilon$: 
		
		$$
		f(\delta v^+)>0 \text{ and } f(\delta v^-)<0.
		$$ 
		Thus if $0$ is a critical point of $f$, it must be a saddle point.
		We prove the statement above by induction on the number of variables $n$ in $f$.
		
		\medskip
		\noindent {\rm \texttt{Basis step}}: 
		Assume $n=1$. If $f(0)=0$ and $f$ is non-constant, then $f=\beta x_1$ where $\beta \neq 0$. 
		By assigning $v^+=(\texttt{sgn}(\beta))$, $v^-=(-\texttt{sgn}(\beta))$ and $\epsilon$ be any positive real number, then, for any $\delta$ such that $0<\delta<\epsilon$, we have:
		\begin{align*}
		    f(\delta v^+) &= |\beta| \cdot \delta >0, \\
		    f(\delta v^-) &= -|\beta| \cdot \delta <0.
		\end{align*}
		
		\medskip
		\noindent {\rm \texttt{Inductive Step}}: Every multilinear function with $n$ $(n\ge 2)$ variables can be decomposed as:
		\begin{align*}
		    f(x_1,\;\dots, \;x_n) = x_1 \cdot g(x_2,\;\dots, \;x_{n}) + h(x_2,\; \dots, \;x_{n}).
		\end{align*}

		Note that both $g$ and $h$ are multilinear functions not depending on $x_1$. Since $f$ is non-constant, without loss of generality, we assume $g(x_2,...,x_n)\not\equiv 0$. 

		Therefore $g\not \equiv 0$. The rest of this proof is provided case by case. 
		Note that $h(0)=0$ because $f(0)=0\cdot g(0)+h(0)=0$.
		\begin{itemize}
			\item $h$ is constant, i.e., $h\equiv 0$. Then $f=x_1g(x_2,...,x_{n})$. 
			\begin{itemize}
				\item $g(x_2,...,x_{n})$ is constant, say $\beta$. We are left with $f=\beta x_1$, a case covered by basis.
				\item $g(x_2,...,x_{n})$ is non-constant.
				\begin{itemize}
					\item[$\square$] If $g(0)=0$, then by inductive hypothesis, $\exists v^+_g$, $v^-_g$ and $\epsilon_g$, such that $g(\delta v^+_g)>0$, $g(\delta v^-_g)<0$ hold for every $0<\delta<\epsilon_g$. Set $v^+=(1,v^+_g)$, $v^-=(1,v^-_g)$ or  and $\epsilon=\epsilon_g$. Now for any $\delta$ such that $0<\delta<\epsilon$,
					$$
					f(\delta v^+)=\delta\cdot g(\delta v^+_g)>0
					$$
					$$
					f(\delta v^-)=\delta\cdot g(\delta v^-_g)<0
					$$
					It is worth mentioning that we could also set  $v^+=(-1,v^-_g)$ and $v^-=(-1,v^+_g)$.
					\item[$\square$] Else if $g(0)\neq 0$, let $v_+=(\texttt{sgn}(g(0)),\;0, \;\dots,\;0)$, $v_-=(-\texttt{sgn}(g(0)),\;0,\;\dots,\;0)$ and $\epsilon$ be any positive number. For any  $\delta$ such that $0<\delta<\epsilon$,
					\begin{align*}
					    f(\delta v^+) &= \delta\cdot \texttt{sgn}(g(0))\cdot g(0)>0, \\
					    f(\delta v^-) &= \delta\cdot (-\texttt{sgn}(g(0)))\cdot g(0)<0.
					\end{align*}
				\end{itemize}
			\end{itemize}
			\item $h$ is non-constant, by inductive hypothesis, $\exists v^+_h$, $v^-_h$ and $\epsilon_h$ such that, $h(\delta v^+_h)>0$, $h(\delta v^-_h)<0$ hold for every $0<\delta<\epsilon_h$. We construct $v^+=(0,v^+_h)$, $v^-=(0,v^-_h)$ and set $\epsilon=\epsilon_h$. For any $\delta$ such that $0<\delta<\epsilon$,
			$$
			f(\delta v^+)=h(\delta v^+_h)>0,
			$$
			$$
			f(\delta v^-)=h(\delta v^-_h)<0.
			$$
		\end{itemize}   
		
		\hfill\qedsymbol
		
\subsection{Proof of Lemma \ref{lemma:local}}
Since $a^\star$ is a  local minimum, then for every $i\in [n]$, either $a^\star_i\in\{\pm 1\}$
		or $\frac{\partial{F_f}}{x_i}\left(a^\star_{[n]-\{i\}}\right) = 0$ holds. Otherwise the gradient would give us a negative direction to decrease the function value. Let $I=\{i \;| \;a^\star_i\in\{\pm 1\}\}$. Consider $F_{I\gets a^\star_{I}}$, where every variable in $I$ is assigned to its value in $a^\star$. We have $a^\star_{[n]-I}$ is a critical point of $F_{I\gets a^\star_I}$ since each  derivative of variable with index in $[n]-I$ is zero.
		
		Suppose $F_{I\gets a^\star_I}$ is non-constant. Since $a^\star_{[n]-I}$ is a critical point inside the cube, by Lemma \ref{lemma:saddle}, $a^\star_{[n]-I}$ is a saddle point of $F_{I\gets a^\star_I}$, which means $a^\star$ is not a local minimum of $F$. Thus $F_{I\gets a^\star_I}$ must be constant and $a^\star$ is feasible follows by Definition \ref{feasible}.
		
		\hfill\qedsymbol

\subsection{Proof of Proposition \ref{projection}}
	\begin{itemize}[leftmargin=0.5cm]
	    \item[--] If $y_i\in [-1,1]$, suppose $\Pi_{[-1,1]^n}(y)_i\neq y_i$, then let $z_i=y_i$ and $z_j=\Pi_{[-1,1]^n}(y)_j$ for $j\neq i$. We have $z\in [-1,1]^n$ and $\|z-y\|_2^2 \le \left\|\Pi_{[-1,1]^n}(y)-y \right\|_2^2$, a contradiction with the definition of the projection.
	    \item[--] If $|y_i|>1$, without loss of generality, suppose $y_i>1$ and $\Pi_{[-1,1]^n}(y)_i<1$. Let $z_i=1$ and $z_j=\Pi_{[-1,1]^n}(y)_j$ for $j\neq i$. Similarly, $z\in [-1,1]^n$ and $\|z-y\|_2^2 \le \left\|\Pi_{[-1,1]^n}(y)-y \right\|_2^2$ still hold. 
	\end{itemize}\hfill\qedsymbol
	
To prove Theorem \ref{theo:convergence_rate}, we need to introduce and prove Proposition \ref{lipschitz}, Proposition \ref{prop:G_and_F} and Lemma \ref{lemma:descent}.
		
\subsection{Proof of Proposition \ref{lipschitz}}
	\begin{enumerate}
	    \item By triangle inequality,
	    \begin{equation}\nonumber
	        \begin{split}
	           |&F(x)-F(y)| \le|F(x_1,...,x_n)-F(y_1,x_2,...,x_n)|
	   +|F(y_1,x_2,...,x_n)-F(y_1,y_2,x_3,...,x_n)|\\
	           &+\cdots+|F(y_1,...y_{n-1},x_n)-F(y_1,...,y_n)| \\
	            =|&(x_1-y_1)\nabla_1 F(x_1,...,x_n)|+\cdots+|(x_n-y_n)\nabla_n F(y_1,...,y_{n-1},x_n)|\\
	        \end{split}
	    \end{equation}
	     Since $\nabla_i F(x)=\frac{\partial F}{\partial x_i}(x)=\frac{1}{2}(F_{i\gets 1}(x)-F_{i\gets -1}(x))\in [-\alpha,\alpha]$ for all $i\in [n]$, we have:
	     \begin{equation}\nonumber
	        \begin{split}
	            &|(x_1-y_1)\nabla_1 F(x_1,...,x_n)|+\cdots+|(x_n-y_n)\nabla_n F(y_1,...,y_{n-1},x_n)|
	            \le \alpha (\sum_{i=1}^n|x_i-y_i|)\\ \le& \alpha n^{\frac{1}{2}}||x-y||_2
	        \end{split}
	    \end{equation}

	    \item Since $\nabla_i F(x)=\frac{\partial F}{\partial x_i}(x)\in [-\alpha,\alpha]$ for all $i\in [n]$, by 1. of Proposition \ref{lipschitz} we have 
	    \begin{equation}\nonumber
	    \begin{split}
	        ||\nabla F(x)-\nabla F(y)||
	        =(\sum_{i=1}^n (\nabla F_i(x)-\nabla F_i(y))^2 )^{\frac{1}{2}}
	        \le(n\cdot (\alpha n^{\frac{1}{2}}||x-y||_2)^2)^\frac{1}{2}
	        =\alpha n ||x-y||_2
	    \end{split}
	    \end{equation}
	    \item Similarly, $\nabla_{i,j}^2F(x)\in [-\alpha,\alpha]$ for all $i,j\in [n]$. By applying 1 of Proposition \ref{lipschitz}. again and the fact that l2 norm is no more than Frobenius norm, we get:
	    \begin{equation}\nonumber
	        \begin{split}
	            ||\nabla^2 F(x)-\nabla^2 F(y)||_2\le & ||\nabla^2 F(x)-\nabla^2 F(y)||_F
	            = \sum_{i,j\in [n]}(\nabla^2_{i,j} F(x)-\nabla^2_{i,j} F(y))^2
	            \\ \le  &(n^2 (\alpha n^{\frac{1}{2}}||x-y||_2)^2)^\frac{1}{2}
	            =  \alpha n^{\frac{3}{2}}||x-y||_2
	        \end{split}
	    \end{equation}
	\end{enumerate}
 \hfill\qedsymbol

 	\begin{proposition} 
	\label{prop:G_and_F} (Inequality of gradient mapping)
	For a function $F$ and every $x\in \texttt{dom}(F)$,
	$$
	 \left \langle \nabla F(x), ~G(x) \right \rangle \ge \|G(x)\|^2_2
	$$
		where $G(x)=\frac{1}{\eta} \left(x-\Pi_{\Delta} \left(x-\eta\nabla F(x) \right) \right)$ is the gradient mapping, and $\Delta$ is a convex set.
	\end{proposition}
 \subsection{Proof of Proposition \ref{prop:G_and_F}}
 
 	Since $x_{t+1}$ is the optimum of the Euclidean projection onto convex set $\Delta$, we have 
	$$
	\langle x_{t+1}-x_{t+1}',z-x_{t+1}\rangle \ge 0,
	$$
	for all $z\in \Delta$. 
	Let $z=x_t$. Notice that $x_{t+1}-x_{t+1}'=G(x_t)-\nabla F(x_t)$ and $x_t-x_{t+1}=-G(x_t)$, we get:
	$$
	   \langle G(x_t)-\nabla F(x_t),-G(x_t)\rangle \ge 0, 
	$$
	and the statement follows directly.\hfill\qedsymbol

	\begin{lemma}\label{lemma:descent} (Descent Lemma). For a multilinear polynomial $F:[-1,1]^n\to [-\alpha,\alpha]$ and its projected gradient mapping $G(\cdot)$, if the step size satisfies $\eta\le \frac{1}{\alpha n}$, then the projected gradient descent (PGD) sequence ${x_t}$ satisfies:
	$$
	    F(x_{t+1})-F(x_t)\le -\tfrac{\eta}{2} \|G(x_t)\|_2^2.
	$$
	
	\end{lemma}

\subsection{Proof of Lemma \ref{lemma:descent}}
	According to the $\left(\alpha n\right)$-gradient Lipschitz continuity property Proposition \ref{lipschitz}, we have:
	\begin{align*}
	    F(x_{t+1}) &\le F(x_t) + \langle\nabla F(x_t),~x_{t+1}-x_t\rangle + \tfrac{\alpha n}{2} \cdot \|x_{t+1}-{x_t} \|^2 \\
	               &= F(x_t)-\eta\langle \nabla F(x_t),~G(x_t)\rangle + \tfrac{\eta^2\alpha n}{2}||G(x_t)||^2 \\
	               &\le F(x_t)-\tfrac{\eta}{2}\|G(x_t)\|^2.
	\end{align*}
	
Note that the last inequality follows by  Proposition \ref{prop:G_and_F}. \hfill\qedsymbol

\subsection{Proof of Theorem \ref{theo:convergence_rate}}
   Assume that $\eta= \frac{1}{nm}$. 
	    Then, the recursion in Lemma \ref{lemma:descent} satisfies:
	    \begin{align*}
	        F(x_{t+1}) &\le F(x_t)-\tfrac{1}{2n m}\|G(x_t)\|^2.
	    \end{align*}
	    Combining all the iterations together, for $T$ iterations, we have:
        \begin{align*}
            F(x_{T+1}) &\leq F(x_T) -\tfrac{1}{2n m}\|G(x_T)\|^2 \\
            F(x_{T}) &\leq F(x_{T-1}) -\tfrac{1}{2n m}\|G(x_{T-1})\|^2 \\
            &\;\;\vdots\\
            F(x_{1}) &\leq F(x_0) -\tfrac{1}{2n m}\|G(x_0)\|^2
        \end{align*}
        Summing all these inequalities, and under the observation that $F(x^\star) \leq F(x_{T+1})$, we get the following:
        \begin{align*}
            \tfrac{1}{2n m} \sum_{t = 0}^T \|G(x_t)\|^2 \leq F(x_0) - F(x^\star).
        \end{align*}
        
        This implies that, even if we continue running gradient descent for many iterations, the sum of gradient norms is always bounded by something constant; this indicates that the gradient norms that we add at the very end of the execution has to be small, which further implies convergence to a stationary point.

        Further, we know:
        $$(T+1) \cdot \min_{t} \|G(x_t)\|^2 \leq \sum_{t = 0}^T \|G(x_t)\|^2.$$
        Then,
        \begin{align*}
            \tfrac{T+1}{2nm} \cdot \min_{t} \| G(x_{t}) \|^{2} &\leq \tfrac{1}{2nm} \sum\limits_{t = 0}^{T} \|G(x_{t})\|^{2} \leq F(x_{0}) - F(x^{\star})\\ \Rightarrow 
            \min_{t} \|G(x_t)\|^2 &\leq \tfrac{2nm}{T+1} \cdot \left(F(x_{0}) - F(x^\star)\right) \\
            \min_{t} \|G(x_t)\| &\leq \sqrt{\tfrac{2nm}{T+1}} \cdot \left(F(x_{0}) - F(x^\star)\right)^{\frac{1}{2}} = O\left(\tfrac{1}{\sqrt{T}}\right).
        \end{align*}
	Note that $|F(x_0)-F(x^*)|\le m$. Thus, to achieve a point such that $\|G(x_T)\| \leq \varepsilon$, we require:
	\begin{align*}
	    \min_{t} \|G(x_t)\| \leq \varepsilon \Rightarrow \sqrt{\tfrac{2nm}{T+1}} \cdot \left(F(x_{0}) - F(x^\star)\right)^{\frac{1}{2}} \leq \varepsilon \Rightarrow O\left(\tfrac{n m^2}{\varepsilon^2}\right) \; \text{iterations}.
	\end{align*} \hfill \qedsymbol

\subsection{Proof of Proposition \ref{prop:localmin-firstorder}}
  For every $i\in[n]$, by definition of gradient mapping, $G(x)_i=0$ means
	    $$
	    x_i=\Pi_{[-1,1]}(x_i-\eta \nabla F(x)_i)
	    $$
	    Since $x$ is feasible, if $x_i\in(-1,1)$. i.e. $i\in [n]-I$, then $\nabla F(x)_i=0$. When $\nabla F(x)_i\neq0$, $x_i\neq x_i-\eta \nabla F(x)_i$. By Proposition \ref{projection}, if $x_i=-1$, then $\nabla F(x)_i>0$; if $x_i=1$, then $\nabla F(x)_i<0$.
	    
	    We need to prove $x$ has the lowest function value among all the points in  intersection of $x$'s neighbourhood $\mathcal{N}_{\delta}(x)$ and cube $[-1,1]^n$. Consider all the directions $v$ ($\norm{v}\neq 0$) where $x$ can move a small step to and $x+\eta\cdot v$ still remain in $\mathcal{N}_{\delta}(x)\cup [-1,1]^n$. We have the following constraints for $v$:
	    \begin{equation}\nonumber
	    \begin{cases}
	        v_i\ge 0, \text{if $x_i=-1$},  \\
	        v_i\le 0, \text{if $x_i=1$}.
	    \end{cases}
	    \end{equation}
	    If $x_i\in(-1,1)$, there is no restriction for $v_i$.
	    \begin{itemize}
	        \item[--] If for all $i\in I$, $v_i=0$. Since $x$ is a feasible  solution of $F$, changing the value of variables with indices only from $[n]-I$ does not change the function value. Thus $F(x)=F(x+\eta \cdot v)$.
	        \item[--] Else there exists $i^*\in I$ s.t. $v_{i^*}\neq 0$.
	    The \textbf{directional derivative} at $x$ in direction $v$, denoted as $\nabla_vF(x)$, can be computed by:
	    $$
	    \nabla_vF(x)=\sum_{i\in[n]}v_i\nabla F(x)_i.
	    $$
	    First, note that $v_i\nabla F(x)_i\ge 0$ for all $i\in[n]$ ,thus $\nabla_vF(x)\ge 0$. 
	     Since $\nabla F(x)_{i^*}\neq 0$ and $v_{i^*}\neq 0$ we have $\nabla_vF(x)>0$ for any $v$  and  $F(x)<F(x+\eta \cdot v)$. 
	        \end{itemize}
	        Therefore, $x$ is a local minimum.
	    \hfill\qedsymbol
	    
\begin{algorithm}[h!]
    \SetAlgoLined
    \SetKwInOut{Input}{Input}
    \SetKwInOut{Output}{Output}
    \Input{Polynomial $F$, non-feasible critical point $x$ of $F$, step size $\eta > 0$.}
    \Output{$x$ after moving towards a negative direction.}
    \vspace{0.1cm}
    \hrule
    \vspace{0.1cm}
      $I\gets\{i\;|\;x_i\in\{-1,1\}\}$\\
      $F_{N}(y):= F_{I\gets x_I}(y+x_{[n]-I})-F(x)$ \hfill //\texttt{Lemma \ref{lemma:saddle}}\\ 
     $v=\texttt{NegDirectionSaddle}(F_N)$ \\
     \Return $x+\eta\cdot v$
 \caption{$\texttt{DecInnerSaddle}(F,\;x, \;\eta)$}
 	\label{algo:DecreaseAtInnerSaddle}
\end{algorithm}
	
\begin{algorithm}[h!]
    \SetAlgoLined
    \SetKwInOut{Input}{Input}
    \SetKwInOut{Output}{Output}
    \Input{Polynomial $F$ such that $F(0)=0$.}
    \Output{A negative direction $v^-$ of $F$ at point $0$.} 
    \vspace{0.1cm}
    \hrule
    \vspace{0.1cm}
     \hfill // \texttt{See proof of Lemma \ref{lemma:saddle}}\\
     \eIf{$\texttt{isZero}(F_{1\gets 0})= \texttt{True}$}{
            \eIf{$\texttt{isZero}(F_{1\gets 1}-F(1,0,...,0))= \texttt{True}$}{
            
                $v^- = (\texttt{sgn}(F_{1\gets 1}),0,...,0)$
            }
            {
                \eIf{$F(1,0,...,0)=0$}{$v^- = (1,\texttt{NegDirectionSaddle}(F(1,0,...,0)))$}
                {$v^- = (-\texttt{sgn}(F(1,0,...,0),0,...,0))$}
            }
        }{
          $v^-\gets(0,\texttt{NegDirectionSaddle}(F_{1\gets 0}))$
        }
        \Return $v^-$
 \caption{$\texttt{NegDirectionSaddle}(F)$}
 	\label{algo:find_negative_direction}
\end{algorithm}

\begin{algorithm}[h!]
    \SetAlgoLined
    \SetKwInOut{Input}{Input}
    \SetKwInOut{Output}{Output}
    \Input{Polynomial $F$}
    \Output{\texttt{True}/\texttt{False} on $F\equiv0$}
       \vspace{0.1cm}
    \hrule
    \vspace{0.1cm}
      $x \sim \mathcal{U}[-1,1]^n$\\
      \Return \texttt{True} if $F(x)=0$, otherwise \texttt{False}
    
 \caption{$\texttt{isZero}(F)$}
 	\label{algo:Constant_Testing}
\end{algorithm}

	\begin{algorithm}[h!]
    \SetAlgoLined
    \SetKwInOut{Input}{Input}
    \SetKwInOut{Output}{Output}
    \Input{Polynomial $F$, feasible critical point $x$.}
    \Output{
     $x\in[-1,1]^n$; \\
    $\texttt{LocalMinFlag}\in\{\texttt{True,False,Unknown}\}$: if $x$ is a local minimum of $F$ in $[-1,1]^n$.
    }
       \vspace{0.1cm}
    \hrule
    \vspace{0.1cm}
     $J = \{j \; | \; \nabla F(x)_j=0\}$ \quad \text{and} \quad 
     $I =  \{i\; | \; x_i\in\{\pm 1\} \wedge i\in J\}$\\
                    $H = \nabla^2 F_{[n]-J\gets x_{[n]-J}}(x_{J})$\\
                    \uIf{$\exists i\in I, ~ j\in J-I$ such that $H_{ij}\neq 0$ \label{line:Hessian_condition1}}
                    {
                        $v\gets (0,0,...,v_i=-\texttt{sgn}(x_i),0,...,0,v_j=\texttt{sgn}(x_i\cdot H_{ij}),0,...,0)$ \hfill // \texttt{see Prop \ref{prop:negative_given_by_Hessian}\label{line:negative_direction}}\label{line:negative_direction1}\\
                        \Return ($x+\eta \cdot v$, \texttt{False})
                    }
                    \uElseIf{$\exists i_1,i_2\in I$ ~ such that $H_{ij}\cdot x_{i_1}x_{i_2}< 0$\label{line:Hessian_condition2}}
                    {
                        $v\gets (0,0,...,v_i=-\texttt{sgn}(x_{i_1}),0,...,0,v_j=-\texttt{sgn}(x_{i_2}),0,...,0)$ \hfill // \texttt{see Prop \ref{prop:negative_given_by_Hessian}\label{line:negative_direction2}}\\
                        \Return ($x+\eta \cdot v$, \texttt{False})
                    } 
                    \uElseIf{$\forall i,j,~ i\neq j$ and at least one of $i$ and $j$ is in $I$  such that $H_{ij} \neq 0 $}
                    {
                        \Return ($x$, \texttt{True}) \hfill // \texttt{See Prop \ref{prop:localmin-secondorder}} \label{line:secondorderlocalmin}
                    }
                    \Else
                    {
                    \Return ($x,\texttt{Unknown}$)
                    //\texttt{We meet a very rare degenerate saddle point at the corner. Do random restart} 
                    }
 \caption{$\texttt{useHessian}(F,x)$}
 	\label{algo:ProcessHessian}
\end{algorithm}

	\begin{proposition}
	\label{prop:negative_given_by_Hessian}
	When algorithm \ref{algo:ProcessHessian} reaches line \ref{line:negative_direction1} or \ref{line:negative_direction2}, $v$ is a negative direction of $F$. I.e., $F(x+\eta v)<F(x)$ and $x+\eta v$ is within the cube $[-1,1]^n$.
	\end{proposition}
	\subsection{Proof of Proposition \ref{prop:negative_given_by_Hessian}}
	Consider partially assigned function $F'=F_{[n]-J\gets x_{[n]-J}}$. By definition of $J$, $\nabla F'(x_J)=\mathbf{0}$. In the following cases, we will use the \textbf{second directional derivative} of $F'$ at $x_J$ in direction $v$, denoted by $\nabla^2F_v'(x_J)$ as our tool, which can be computed as:
	$$\nabla^2F_v'=\sum_{i,j\in J}v_iv_j H_{ij}=2\sum_{i,j\in J,i\neq j}v_iv_j H_{ij},
	$$
	where $H$ is the Hessian of $F'$ at $x_J$ and the last step follows by multilinearity of $F'$ ($H_{ii}=0$ for every $i\in[n]$).
	\begin{itemize}
	    \item[--] Suppose Algorithm \ref{algo:ProcessHessian} reaches line \ref{line:negative_direction1}. First note that given $\eta$ being small enough, $x_J+\eta \cdot v\in[-1,1]^{|J|}$ because $x_i\in\{\pm 1\}$ and $x_i-\eta\cdot\texttt{sgn}(x_i)\in[-1,1]$; $x_j\in(-1,1)$ thus $x_j+\texttt{sgn}(x_i\cdot H_{ij})\in [-1,1]$. Moreover, $$
	    \nabla^2F_v'=2\cdot (-\texttt{sgn}(x_i)) \texttt{sgn}(x_i\cdot H_{ij})\cdot H_{ij}<0
	    $$
	    Thus, moving towards $v$ will decrease $\nabla F'_v$. Since $\nabla F'(x_J)=\mathbf{0}$ we have $\nabla F'_v(x_J)=0$. Therefore $\nabla F'_v(x_J+\eta v)<0$ and $F'(x_J+v)<F'(x_J)$. Hence $v$ is a also negative direction of $F$ at $x$ since $v$ keeps coordinates of $x$ in $[n]-J$ unchanged. 
	    \item[--] Suppose Algorithm  \ref{algo:ProcessHessian} reaches line \ref{line:negative_direction2}. Also it is easy to confirm that $x_J+\eta \cdot v\in [-1,1]^{|J|}$. Meanwhile,
	    $$
	    \nabla^2F_v'=2\cdot (-\texttt{sgn}(x_{i_1})) (-\texttt{sgn}(x_{i_2}))\cdot H_{ij}<0
	    $$
	    Note that the inequality follows by the condition in line \ref{line:Hessian_condition2} of Algorithm \ref{algo:ProcessHessian}. Similar analysis with the case above guarantees $v$ is a negative direction. 
	\end{itemize}
	
	\begin{proposition}
	\label{prop:localmin-secondorder}
	When algorithm \ref{algo:ProcessHessian} reaches line \ref{line:secondorderlocalmin}, $x$ is a local minimum.
	\end{proposition}
	\subsection{Proof of Proposition \ref{prop:localmin-secondorder}}
	Similar with proof of Proposition \ref{prop:negative_given_by_Hessian}, let $F'=F_{[n]-J\gets x_{[n]-J}}$. We have $\nabla F'(x_J)=\mathbf{0}$.
        Recall that
        $$\nabla^2F_v'=2\sum_{i,j\in J,i\neq j}v_iv_j H_{ij}
	$$
	Consider all the directions $v$ ($\norm{v}\neq 0$) where $x+\eta \cdot v\in[-1,1]^{|J|}$. When the algorithm runs line \ref{line:secondorderlocalmin}, neither of condition in line \ref{line:Hessian_condition1} nor \ref{line:Hessian_condition2} holds. Also $H_{ij}=0$ for every $i,j\in J-I$ because $x_J$ is a   feasible solution of $F'$ so that $F'_{I\gets x_I}$ is constant. Therefore $v_iv_j H_{ij}\ge 0$ for all $i,j\in J$ and
	    $\nabla^2F_v'=2\sum_{i,j\in J,i\neq j}v_iv_j H_{ij}\ge 0$.
	
	    \begin{itemize}
	        \item[--] If $v$ has two or more non-zero elements then $\nabla^2F'_v>0$. Since $\nabla F'_v=0$, $F'(x_J)< F'(x_J+\eta v)$.
	     \item[--]Else If $v$ has only one non-zero element, say $v_1\neq 0$, then $\nabla^2F'_v(x_J+\eta\cdot v)=H_{11}=0$. Thus moving towards $v$ does not change $\nabla F'_v$, combining which with $\nabla F_1'(x_J)=0$ we know that moving towards $v$ does not change $F'$, i.e., $F'(x_J)= F'(x_J+\eta v)$.
	     \end{itemize}
	     Therefore $x_J$ is a local minimum of $F'$ and we will use it  to show $x$ is a local minimum of $F$. Suppose $x$ is not a local minimum of $F$ and there exists $v$ s.t. $x+\eta \cdot v\in[-1,1]^n$ and $F(x+\eta \cdot v)<F(x)$. Then there exists $i\in [n]-J$ s.t. $v_i\neq 0$, otherwise $x_J$ is not a local  minimum of $F'$. However, recall $G(x)=\mathbf{0}$ and $\nabla F(x)_i\neq 0$ because $i\in[n]-J$, the directional derivative $\nabla F_v(x)>0$ by similar analysis in the proof of Proposition \ref{prop:localmin-firstorder}, which conflicts with $F(x+\eta \cdot v)<F(x)$.\hfill\qedsymbol
	     
		In Algorithm \ref{algo:main_algorithm}, we used Algorithm \ref{algo:Constant_Testing} to test whether a multilinear polynomial is equivalent to 0 probabilistically.  In this way, we bypass tedious representations of multilinear polynomials. Proposition \ref{prop:constant_testing} shows this testing  method is always correct in the sense of probability.
	\begin{proposition} (Constant Testing)\label{prop:constant_testing}	For a  multilinear polynomial $F\not\equiv0$ and a uniformly random vector $x$ from $[-1,1]^n$, $$\mathop{\mathbb{P}}\limits_{x\sim [-1,1]^n}[F(x)=0]=0$$ 
	\end{proposition}

\subsection{Proof of Proposition \ref{prop:constant_testing}}
	If $F$ is a constant other than 0, the statement holds trivially. Thus we assume $F$ is non-constant.
	
	\texttt{Basis}: $n=1$. $F(x)=ax_1+b$ where $a\neq 0$. $F(x)=0$ has only one root $x^*_1=-\frac{b}{a}$. Thus $\mathop{\mathbb{P}}\limits_{x\sim [-1,1]^n}[F(x)=0]=0$.
	
	\texttt{Inductive Step}: suppose $F$ has $n$ variables ($n\ge 2$). Decompose $F$ as:
	$$
	F(x)=x_1g(x_2,...,x_n)+h(x_2,...,x_n)
	$$
	\begin{itemize}
	\item If both $h$ and $g$ are constant, the case degenerates to the basis.
	\item If $g(x)$ is  constant and $h(x)$ is non-constant, let $g(x)\equiv a$. Then we assume $a\neq 0$, otherwise $F(x)$ degenerates to a polynomial with $n-1$ variables.
	 \begin{equation}\nonumber
	    \begin{split}
	        \mathop{\mathbb{P}}\limits_{x\sim [-1,1]^n}[F(x)=0]=\mathop{\mathbb{P}}\limits_{x_1\sim [-1,1]}[x_1=-\frac{h(x)}{a}]=0
	    \end{split}
	\end{equation} 
	
	\item If $g(x)$ is non-constant and $h(x)$ is constant. Let $h(x) \equiv b$\\
	\begin{itemize}
	    \item If $b=0$,
	    \begin{equation}\nonumber
	    \begin{split}
	        &\mathop{\mathbb{P}}\limits_{x\sim [-1,1]^n}[F(x)=0]
	        \le\mathop{\mathbb{P}}\limits_{x\sim [-1,1]^{n-1}}[g(x)=0]+\mathop{\mathbb{P}}\limits_{x_1\sim [-1,1]}[x_1=0]=0
	    \end{split}
	\end{equation} 
	    \item If $b\neq 0$,
	 \begin{equation}\nonumber
	    \begin{split}
	        &\mathop{\mathbb{P}}\limits_{x\sim [-1,1]^n}[F(x)=0]
	        =\mathop{\mathbb{P}}\limits_{x\sim [-1,1]^n}[g(x)\neq 0]\cdot\mathop{\mathbb{P}}\limits_{x_1\sim [-1,1]}[x_1=-\frac{h(x)}{g(x)}]=0
	    \end{split}
	\end{equation} \end{itemize}
	
	\item If both $h$ and $g$ are non-constant:
	    \begin{equation}\nonumber
	    \begin{split}
	        &\mathop{\mathbb{P}}\limits_{x\sim [-1,1]^n}[F(x)=0]
	        =\mathop{\mathbb{P}}\limits_{x\sim [-1,1]^{n-1}}[g(x)=0]\cdot\mathop{\mathbb{P}}\limits_{x\sim [-1,1]^{n-1}} [h(x)=0]\\&+\mathop{\mathbb{P}}\limits_{x\sim [-1,1]^n}[g(x)\neq 0]\cdot\mathop{\mathbb{P}}\limits_{x_1\sim [-1,1]}[x_1=-\frac{h(x)}{g(x)}]=0
	    \end{split}
	\end{equation} 
	\end{itemize}
	\hfill\qedsymbol
\end{document}